\documentclass[a4paper,11pt]{article}
\pdfoutput=1 
\usepackage[table,dvipsnames]{xcolor}
\usepackage{jinstpub} 
\usepackage{lineno}
\usepackage{multirow}
\usepackage{comment}
\usepackage{graphicx}
\usepackage{tikz}
\usepackage[export]{adjustbox}
\usepackage{siunitx}
\usepackage{booktabs}
\usepackage{float}
\usepackage{tabularx}
\usepackage{placeins}
\usepackage{upgreek}
\usepackage{makecell} 
\usetikzlibrary{decorations.pathmorphing}
\usetikzlibrary{arrows.meta}
\definecolor{mutedgold}{rgb}{0.93, 0.93, 0.93}

\usepackage{colortbl}

\title{\boldmath Characterization of Lateral Amorphous Selenium Photodetectors for Low-Photon and VUV Detection at Cryogenic Temperatures}

\author[a,b,1]{M.~Rooks \note{Corresponding author.}}
\author[c]{S.~Abbaszadeh}
\author[a]{J.~Asaadi,}
\author[a]{V.~A.~Chirayath}
\author[b,2]{M.~Febbraro \note{Current address: Air Force Institute of Technology, WPAFB, OH, USA.}}
\author[d]{M.~Á.~García-Peris}
\author[d]{E.~Gramellini}
\author[c]{K.~Hellier}
\author[e]{B.~Sudarsan}
\author[a]{I.~Tzoka}

\affiliation[a]{University of Texas at Arlington,\\Physics Department Arlington, TX 76019, USA}
\affiliation[b]{Oak Ridge National Laboratory,\\ Physics Division, Oak Ridge, TN 37831, USA}

\affiliation[c]{University of California Santa Cruz \\
Electrical and Computer Engineering, Santa Cruz, CA 95064, USA}

\affiliation[d]{University of Manchester, \\Department of Physics and Astronomy, University of Manchester, Manchester M13 9PL, UK}

\affiliation[e]{Rutgers University, \\Department of Physics and Astronomy, Rutgers University, New Brunswick, NJ 08903, USA}

\emailAdd{michael.rooks@mavs.uta.edu }

\abstract{
The performance of amorphous selenium (a-Se) as a cryogenic photodetector material is evaluated through a series of experiments using laterally structured devices operated in a custom optical test stand. These studies investigate the response of a-Se detectors to low-photon fluxes at high electric fields near avalanche conditions, the linearity of the photoconductive response over a wide dynamic range and the direct detection of narrowband 130\,nm vacuum ultraviolet (VUV) illumination. At 87\,K, matched-filter analysis shows reliable single-shot detection with efficiencies $\geq$\,80\% and area under the curve (AUC) $\geq$\,0.85 using as few as $\sim$6800 incident 401\,nm photons, corresponding to $\sim$3400 photons within field-active regions after accounting for geometric constraints. Measurements are performed at cryogenic temperatures using calibrated photon fluxes derived from a silicon photomultiplier reference and a characterized optical filter stack. Additional experiments using a tellurium-doped a-Se (a-SeTe) device explore the material’s behavior under identical test conditions and demonstrate that avalanche is achievable in a-SeTe at cryogenic temperatures. The results demonstrate reproducible low-noise operation, VUV sensitivity and field-dependent gain behavior in a lateral a-Se architecture, representing the first reported observation of avalanche multiplication in laterally structured a-Se and a-SeTe devices at cryogenic temperatures. These findings support the potential integration of laterally structured a-Se devices into next-generation pixelated liquid-argon time projection chambers (TPCs) requiring scalable, high-field-compatible photon detection systems.
}

\keywords{Cryogenic detectors, Photon detectors, Noble liquid detectors}

\begin{document}

\maketitle
\flushbottom

\section{Introduction}
\label{sec:Intro}

Scintillation light detection in noble liquid time projection chambers (TPCs) is essential for establishing the interaction time $t_0$, which anchors the three-dimensional reconstruction of particle tracks. The light signal also contributes to calorimetric measurements, as the total energy deposited by a particle is shared between scintillation light and ionization charge in a complementary, anticorrelated manner~\cite{Foreman2020}. In addition, the scintillation signal enables particle identification through pulse shape discrimination~\cite{Wahl_2014, Akerib_2018}. These capabilities are particularly important in rare event searches and neutrino experiments, where precise spatial, temporal and calorimetric reconstruction is critical. The increasing interest in pixelated TPC architectures featuring fully segmented anode planes places new demands on photon detection systems. In the pursuit of scalable solutions, integrating photodetection with the charge readout provides a practical means to enhance light collection at the anode and improve the uniformity of photon detection.

Conventional photodetectors such as photomultiplier tubes (PMTs) and silicon photomultipliers (SiPMs) have been widely used in noble liquid detectors \cite{Bonivento2024}. However, scaling these technologies to instrument large areas remains costly, especially when accounting for associated readout electronics and cryogenic compatibility. Their physical footprint, tiling constraints and limited native sensitivity to the vacuum ultraviolet (VUV) light produced by noble liquids often require wavelength shifters~\cite{instruments5010004}, complicating detector integration. These limitations motivate the development of alternative technologies capable of direct VUV detection at low photon flux under cryogenic conditions.

Amorphous selenium is a photoconductive material with a long history in x-ray imaging and radiation detection~\cite{Wronski2008-zf, Zhao2008-tf}. It exhibits strong absorption in the VUV, low dark current and can be thermally evaporated over large areas at low temperatures~\cite{Abbaszadeh2013_1, Abbaszadeh2013_2, Leiga1968, Huda2022}. These properties make a-Se a promising candidate for cryogenic photon detection. Recent studies have demonstrated the structural and electrical stability of a-Se devices in liquid argon environments, with no detrimental impact on argon purity, demonstrating their compatibility with noble liquid TPC operation ~\cite{Rooks2023}.

\begin{figure}[ht]
\centering
\includegraphics[width=\textwidth]{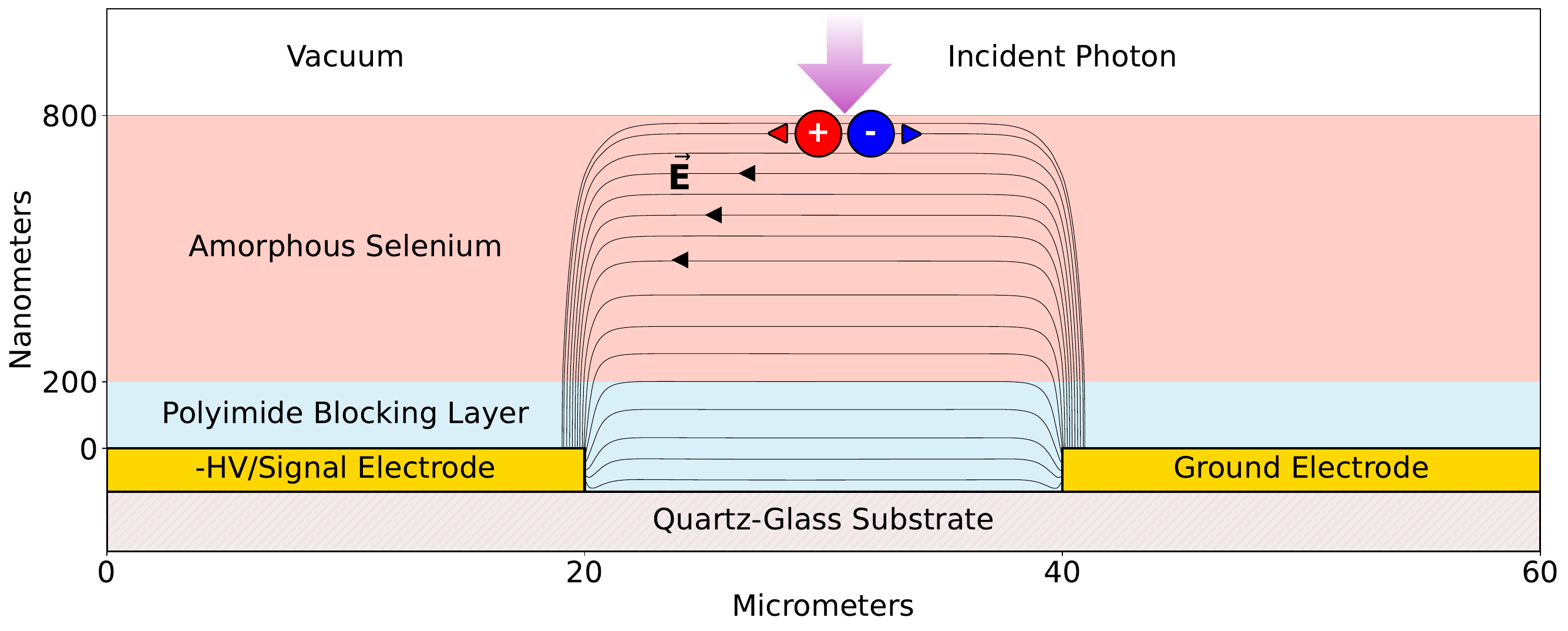}
\caption{Schematic of a lateral amorphous selenium photodetector showing electric field lines simulated using COMSOL Multiphysics\textsuperscript{\textregistered} software with an applied bias of 2~kV corresponding to an electric field of 100~V/\textmu m~\cite{comsol}. For visual clarity, field lines are shown only within the regions active for charge transport and collection. Simulations use relative permittivities of 6.6 for amorphous selenium and 2.9 for polyimide. Incident photons generate electron–hole pairs and under the applied bias, holes drift toward the signal electrode to produce the photocurrent signal.}

\label{fig:lateral_sketch}
\end{figure}

Amorphous selenium generates electron–hole pairs upon absorption of optical or VUV photons. Due to its higher mobility–lifetime product, hole transport dominates the signal formation~\cite{Kasap2011}. At room temperature, hole mobilities typically range from 0.13 to 0.14~cm$^2$/V$\cdot$s, while electron mobilities are significantly lower ($\sim$0.003~cm$^2$/V$\cdot$s)~\cite{Abbaszadeh2013,Kasap2009}. At high electric fields, hole mobility increases due to field-assisted detrapping. However, at cryogenic temperatures, thermal detrapping is suppressed, leading to reduced mobility and lower collected signal~\cite{Hijazi2015}. This behavior has been confirmed in the lateral geometry used in this work in Section \ref{sec:Te_aSe_comparison}.

The detector structure shown in Fig.~\ref{fig:lateral_sketch} uses an interdigitated electrode (IDE) geometry to apply a lateral electric field across the a-Se surface. In this configuration, photogenerated carriers drift parallel to the film surface, confined to the shallow region between opposing electrodes. The electrode gap length $L$ sets the lateral field magnitude $E$ as
\begin{equation}
E = \frac{V}{L},
\end{equation}
where \( V \) is the applied bias. This drift path governs both the carrier transit time and in the high-field regime, the achievable avalanche gain. At fields exceeding the ionization threshold, holes can initiate impact ionization. A longer drift path supports more multiplication events, analogous to the role of thickness in vertical avalanche structures~\cite{Reznik2008}.

The vertical thickness of the a-Se film defines the absorbing volume but does not directly influence the drift distance in lateral devices. Due to the strong absorption of photons with wavelengths below 500 nm in a-Se, most carriers are generated within the top 50~nm of the film~\cite{Leiga1968}, emphasizing the role of surface properties in determining signal formation. Studies have shown that surface trapping and composition can alter local electric fields and impair signal stability~\cite{Abbaszadeh2013, Kasap2022}. In particular, arsenic-rich surface layers formed during deposition have been linked to reduced hole transport. Additionally, Wang et al. reported that enhanced responsivity in thin lateral devices may result from the narrowing of a surface space charge region (SCR) under illumination, effectively increasing the conductive volume~\cite{Wang2010}. While that analysis focused on an a-Se–air interface, similar field redistribution may occur at a-Se–vacuum or a-Se–liquid noble interfaces, depending on surface state density and band alignment.

To enable high-field operation and suppress injection currents, a blocking layer is introduced between the a-Se and the electrodes. In this work, a single polyimide (PI) film is spin-coated over both contacts, forming a continuous injection barrier \cite{Abbaszadeh2012_1, Abbaszadeh2012_2}. This approach simplifies fabrication and has been shown to suppress dark current while allowing stable operation at high lateral fields~\cite{Abbaszadeh2013}. Other approaches have used ultrathin dielectric blocking layers such as ALD-deposited HfO$_2$ or Al$_2$O$_3$ to reduce injection via field redistribution and tunneling control~\cite{Chang2016}. These designs achieve excellent dark current suppression, but present fabrication challenges in conformal coverage and large-area integration.

The present work consolidates several experiments using laterally structured a-Se photodetectors operated at cryogenic temperatures and exposed to low photon flux. These include a threshold study near the avalanche regime under calibrated single-shot excitation, a linearity scan over several orders of magnitude in photon yield at 87~K and a demonstration of direct sensitivity to 130~nm light from an argon flashlamp. For laser-based measurements, photon flux calibration was performed using a SiPM reference and a characterized absorptive filter stack. A comparative data set using a-SeTe devices is also presented to assess material effects and validate the robustness of the lateral architecture. The undoped measurements establish the baseline cryogenic behavior against which the effects of Te-doping are evaluated.

\section{Device Architecture and Fabrication}\label{sec:fabrication}
A lateral device geometry was selected to facilitate direct optical access to the photoconductive layer without the need for a transmissive top electrode. While much of the present study focuses on characterization with 401 nm laser excitation, additional measurements were performed using narrowband 130 nm light to directly assess VUV sensitivity. The design targets application in noble liquid detectors where scintillation light is produced in the VUV, with argon and xenon emitting at 128 nm and 178 nm respectively. In such environments, conventional vertical geometries typically require a VUV-transmissive electrode, which can introduce significant material and fabrication constraints. The lateral configuration avoids this limitation, providing a more flexible and scalable path toward direct VUV-sensitive photodetectors \cite{Rooks2023}. The following sections describe the fabrication and experimental setup used to characterize these devices across a range of excitation conditions and temperatures.

\begin{figure}
\centering
\scalebox{.8}{
\begin{tikzpicture}
  \begin{scope}
    \clip (-7.5,-7) rectangle (7,6); 
    \node[inner sep=0pt] (img) {\includegraphics[width=\textwidth]{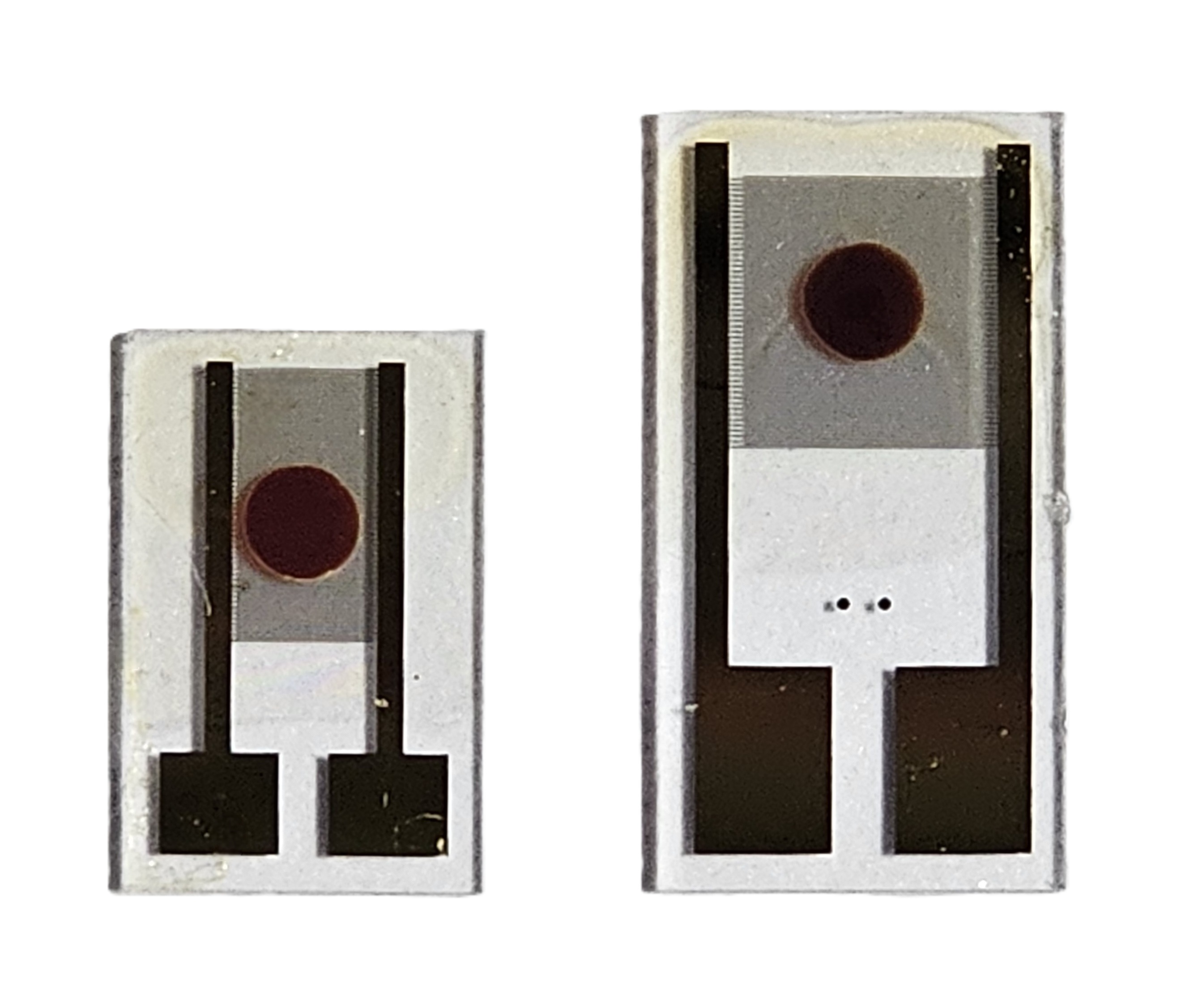}};
  \end{scope}
   
  \draw[<->, black, thick] (-6.0, -5.3) -- (-1.4, -5.3) node[midway, below] {5.2 mm};
  \draw[<->, black, thick] (.9, -5.3) -- (6.1, -5.3) node[midway, below] {5.9 mm};
  
  \draw[<->, black, thick] (-6.5, 2.2) -- (-6.5, -5)
  node[midway, rotate=90, anchor=center, yshift=8pt] {8.2 mm};

  \draw[<->, black, thick] (0.4, 5) -- (0.4, -4.9)
  node[midway, rotate=90, anchor=center, yshift=8pt] {11.5 mm};

  \node at (-3.8,-6.2) {IDE-A};
  \node at (3.5,-6.2) {IDE-B};
  
  \node[text=black] at (-3.7, -2.4) {Polyimide};
  \node[text=black] at (3.5, 0) {Polyimide};
  
  \node[text=white] at (3.5, 2.6) {a-Se};
  \node[text=white] at (-3.7, -0.25) {a-Se};

  \node[text=black] at (-3.6, 3) {IDE Fingers};  
  \begin{scope}[shift={(0.06,-0.2)}]
  \filldraw[fill=mutedgold, draw=black, thick]
  (-3.8,2.6) -- (-3.6,2.6) -- (-3.6,1.9) -- (-3.5,1.9) -- (-3.7,1.7) -- (-3.9,1.9) -- (-3.8,1.9) -- (-3.8,2.6) -- cycle;
  \end{scope} 
  
  \node[text=black] at (3.4, 5.5) {IDE Fingers}; 
  \begin{scope}[shift={(3.4,4.86)}]
    \filldraw[fill=mutedgold, draw=black, thick]
    (0,0) -- (0.2,0) -- (0.2,-0.7) -- (0.3,-0.7) -- (0.1,-0.9) -- (-0.1,-0.7) -- (0,-0.7) -- (0,0) -- cycle;
    \end{scope}  

\end{tikzpicture}
}
\caption{Photograph of the fabricated detectors: IDE-A (left) and IDE-B (right). Overall substrate dimensions are indicated for each device. Both detectors have a centrally located 1.6 mm diameter, 600 nm thick a-Se dot, thermally evaporated onto a 200 nm spin-coated polyimide layer that spans the entire interdigitated region of 50 finger pairs. The field-active region of a single finger pair in IDE-A is illustrated in Fig.~\ref{fig:lateral_sketch}.}
\label{fig:dots}
\end{figure}
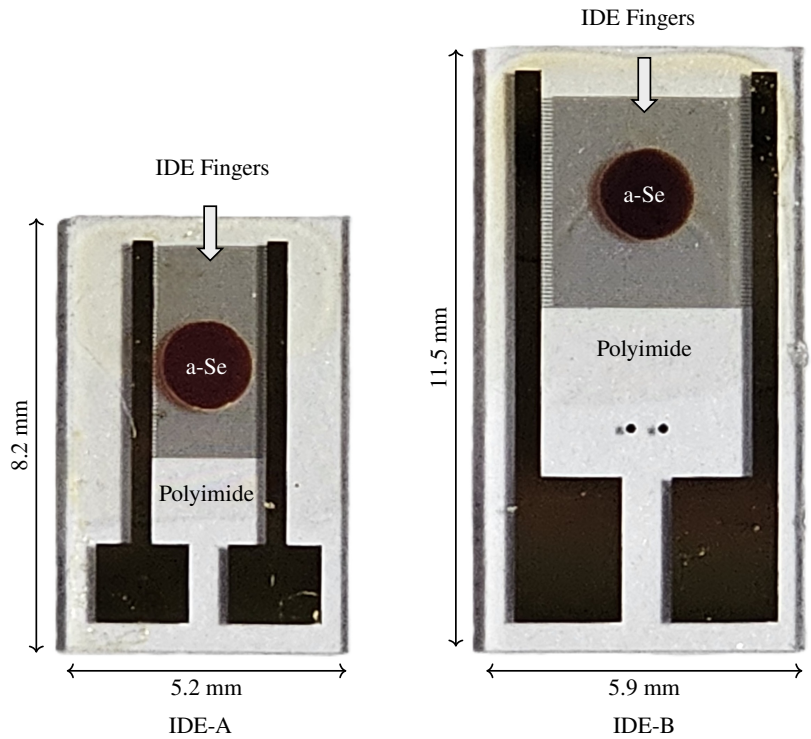

Two detector geometries were used in this study, referred to as IDE-A and IDE-B, where the naming reflects the layout of the IDE on each substrate. Fully fabricated detectors of both designs are shown in Figure~\ref{fig:dots}. IDE-A was used in earlier experiments but became unavailable due to changes in vendor availability. IDE-B was adopted for subsequent measurements and retains the same fabrication process, material stack and functional architecture, with slight differences in electrode layout and substrate dimensions. 

Photodetectors were fabricated using commercially available micro-electro-mechanical systems (MEMS) chips purchased in batch quantities. Each IDE-A unit consisted of a $0.53 \pm 0.01\,\text{mm}$ thick quartz glass substrate with lateral dimensions of $8.21 \pm 0.01\,\text{mm} \times 5.22 \pm 0.01\,\text{mm} $. The electrode structure included 50 pairs of fingers patterned on the substrate using a 30\,nm titanium adhesion layer followed by a 100\,nm gold conductive layer. Each finger was $19.49 \pm 0.06\,\upmu\text{m}$ wide and $2099\pm 0.16\,\upmu\text{m}$ long, with a $20.56\pm 0.23\,\upmu\text{m}$ gap between opposing electrodes.

Each IDE-B unit used a $0.54 \pm 0.01\,\text{mm}$ thick quartz substrate measuring $11.46 \pm 0.02\,\text{mm} \times 5.94 \pm 0.03\,\text{mm} $. The IDE consisted of 50 finger pairs with the same metal stack as IDE-A. Finger widths were $16.22 \pm 0.39\,\upmu\text{m}$, lengths $3070 \pm 0.51\,\upmu\text{m}$ and inter-electrode gaps $22.96 \pm 0.43\,\upmu\text{m}$. Substrate measurements were made with precision calipers, while dimensional measurements of the IDEs were performed using a Keyence VHX-series high-resolution digital microscope. A mechanical profilometer measurement placed the total metal thickness at 130 ± 2 nm for both IDEs. 

Detectors were fabricated in batches of four. Substrates were cleaned using a standard solvent procedure involving sequential rinses in acetone, isopropanol and deionized water, followed by drying with nitrogen gas. Cleaning was performed within an ISO 5 (Class 100) laminar flow hood.

To suppress dark current by inhibiting charge injection from the electrodes and to support stable high-field operation, a thin polyimide layer was deposited over the IDE. PI-2610 polyimide purchased from HD Microsystems was selected for its high dielectric breakdown strength of 200~V/\(\upmu\)m \cite{pi2610}. The polyimide solution was prepared by mixing 2~g of PI-2610 with 1~mL of N-methyl-2-pyrrolidone (NMP, $\geq$99\%) and stored at \(-30\,^{\circ}\text{C}\)~\cite{NMP_Sigma}. Prior to use, the solution was brought to room temperature.

Polyimide was applied by spin-coating the IDE surface with contact pads masked using Kapton tape. The spin process began with a brief dwell time of 20 seconds followed by spinning at 500 rpm for 5 seconds and then 1450 rpm for 180 seconds. The samples were soft-baked on a hotplate at 90°C for 15 minutes, ramped to 150°C at 7°C/min then to 350°C at 4°C/min, cured at 350°C for 30 minutes and cooled to room temperature at 4°C/min. The final film thickness was measured to be 201 nm ± 2 nm using a mechanical profilometer.

Following polyimide deposition, a-Se was thermally evaporated onto the substrate using a retrofitted NRC/Varian 3117 Thermal Evaporator. A molybdenum boat was loaded with 25 mg of stabilized selenium (99.995\% Se, doped with 0.2\% As and 10 ppm Cl) \cite{retorte}. For the tellurium-doped detectors an identical thermal evaporation procedure was followed using selenium pellets containing 10\% Te by weight from the same vendor. The a-SeTe devices share the same fabrication process, polyimide blocking layer and a-Se dot geometry, enabling direct comparison with stabilized-a-Se samples. 

A custom stainless steel shadow mask was used during evaporation to define the active region. The mask contained four circular openings, each 1.6 mm in diameter, positioned to center an a-Se dot over each IDE. The spot diameter was chosen to be slightly larger than the 1.2 mm laser beam profile to ensure full capture of the incident light while minimizing dark current from peripheral selenium coverage. The evaporation rate was maintained at 100 Å/s, monitored using a quartz crystal and an Inficon XTM/2 Deposition Monitor and halted at a total deposition of 10 kÅ. After deposition, the chamber was backfilled to a partial pressure and the samples were held under static vacuum in darkness for 24 hours. The a-Se layer thickness was measured to be 604 ± 2 nm by mechanical profilometry and the circular dot diameter was measured to be 1.62 mm via digital microsope. Samples were stored in a vacuum sealed light-tight container at room temperature when not in use.

\section{Experimental Methods}\label{sec:methods}

\begin{figure}
    \centering
    \includegraphics[width=\textwidth, trim=3.0cm 0 0 0, clip]{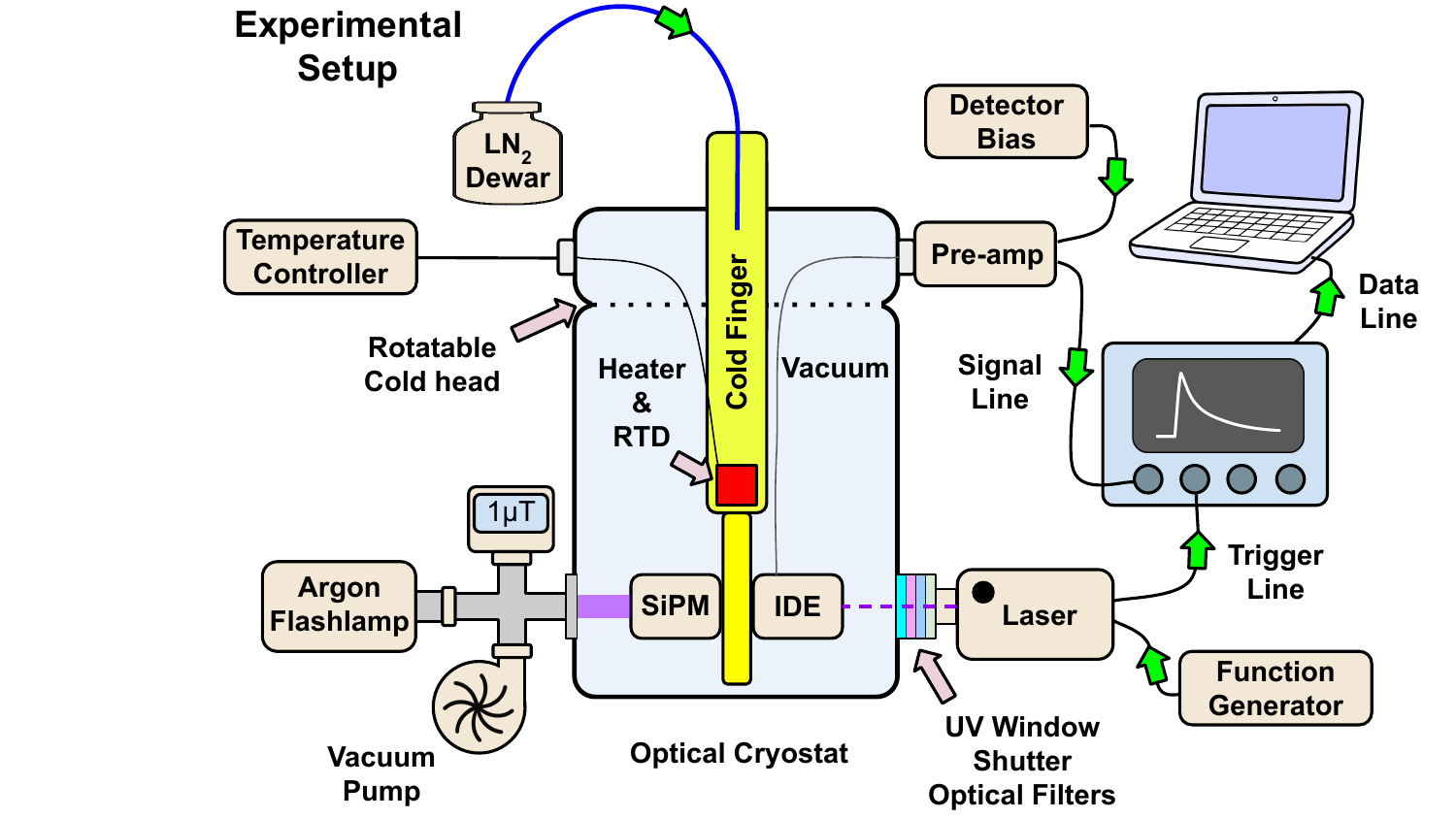}
    \caption{Schematic of the cryogenic optical test system. The optical cryostat features a rotatable liquid nitrogen–cooled cold head with a sample mount that holds the a-Se detector on one side and a SiPM reference on the other. Two optical viewports admit light from a picosecond pulsed laser and an argon flashlamp. Rotating the cold head positions the a-Se detector in the path of the selected light source for optical excitation}
    \label{fig:IDE}
\end{figure}

The structure of the experimental program is outlined below. Figure~\ref{fig:IDE} provides a schematic overview of the cryogenic optical test system, showing the key components used throughout Sections~\ref{sec:cryostat}–\ref{sec:datataking}. Section \ref{sec:cryostat} introduces the cryogenic optical test system, describing the vacuum cryostat, rotatable cold head and surface-temperature mapping used to verify thermal equilibrium at 87 K and 93 K. Section \ref{sec:laser} details the 401 nm picosecond-laser illumination path, neutral-density filter stacks and the SiPM-based photon-flux calibration that anchors all laser-based measurements. The complete signal chain is presented in Section \ref{sec:readout}, along with grounding and noise-mitigation strategies necessary for detecting sub-millivolt signals under cryogenic conditions. Section \ref{sec:ArLamp} describes the modified argon flash-lamp used to deliver narrow-band 130 nm VUV pulses, including the in-vacuum filter housing that preserves spectral purity. Section \ref{sec:datataking} concludes with a summary of the data-taking protocols for low-photon threshold scans, high-field response curves, VUV exposure tests and tellurium-doped comparisons.

\subsection{Cryogenic Optical Test System}
\label{sec:cryostat}

The experimental test apparatus consists of a Lake Shore SuperTran ST-100 Series continuous-flow cryostat with the sample maintained under vacuum. Cooling was achieved using liquid nitrogen supplied from a small external Dewar. A resistive thermal device (RTD) and integrated heater installed in the coldfinger provided closed-loop temperature control with a stability of 10\,mK down to 77.77\,K. Temperature was monitored continuously using a digital readout system and data collection was initiated only after thermal equilibrium was established.

A custom copper sample holder was machined to optimize thermal contact with the detector substrate while maintaining electrical isolation. A thin sheet of high-purity indium foil was placed between the substrate and copper block to improve thermal conductivity across the interface. The holder was designed with two opposing mounting surfaces, one for the a-Se photodetector and the other for a reference SiPM. During standard laser-based measurements, the a-Se device faced an optical viewport equipped with the laser illumination path. For SiPM-based photon flux calibration, the entire coldhead assembly, including the cold finger and sample holder was rotated 180 degrees to position the SiPM in the same optical path. The SiPM was mounted such that its active area occupied the same physical plane as the a-Se detector ensuring consistent beam alignment and illumination geometry between measurements. Electrical connection to the SiPM was provided by a dedicated coaxial cable routed to a separate vacuum feedthrough terminated with a BNC connector.

For the VUV flashlamp experiment, described in Section~\ref{sec:ArLamp} the coldhead was rotated to position the a-Se detector in front of a different viewport coupled to the vacuum system. This port served dual purpose as both the vacuum line and a beamline for the argon flashlamp. The flashlamp output was directed into the cryostat through this port which was connected to a four-way vacuum cross, fitted with an ion gauge and sealed optical path, allowing VUV light to reach the detector surface without atmospheric attenuation.

Laser illumination was delivered through a UV-grade fused silica window with 91.5\% transmission at 401\,nm. The window was coupled to an optical train consisting of a mechanical shutter, lens tubes and neutral density filters. A picosecond-pulsed 401\,nm diode laser was aligned to the optical axis and directed through the window onto the active area of the detector. The laser was mounted to a precision alignment rig equipped with translational controls for horizontal and vertical positioning. The focal point was set at the detector surface with a working distance of approximately 20\,cm and an alignment procedure was performed prior to each run to ensure reproducibility.

The remaining cryostat viewports were sealed with aluminum blanks to ensure the vessel remained light-tight and shielded from stray electromagnetic interference.

\subsubsection{Surface temperature mapping and thermal contact analysis}
To verify the actual detector temperature during cryogenic operation, a surface temperature characterization study was performed for both IDE-A and IDE-B. Since key measurements were targeted at 87\,K for liquid argon, it was necessary to confirm that the detector surface could reach and remain near these conditions during illumination.

Each detector was installed in the cryostat under standard operating conditions. A miniature 100\,$\Omega$ platinum RTD (PT100) was affixed to the surface of the a-Se layer. The sample holder was cooled slowly from room temperature to 77.77\,K and temperature readings were recorded in parallel: one from the cryostat’s base sensor and one from the detector surface via passive ohmmeter readout of the RTD. Measurements were taken at approximately 8--9\,K intervals based on the sample holder temperature. To evaluate surface temperature stability, the system was held at 200\,K, 165\,K, and 77.77\,K for 30 minutes at each point. Across all three cases, the detector surface temperature varied by no more than 10 mK. Passive RTD measurement was used throughout to eliminate self-heating.

\begin{figure}
    \centering
    \includegraphics[width=\linewidth]{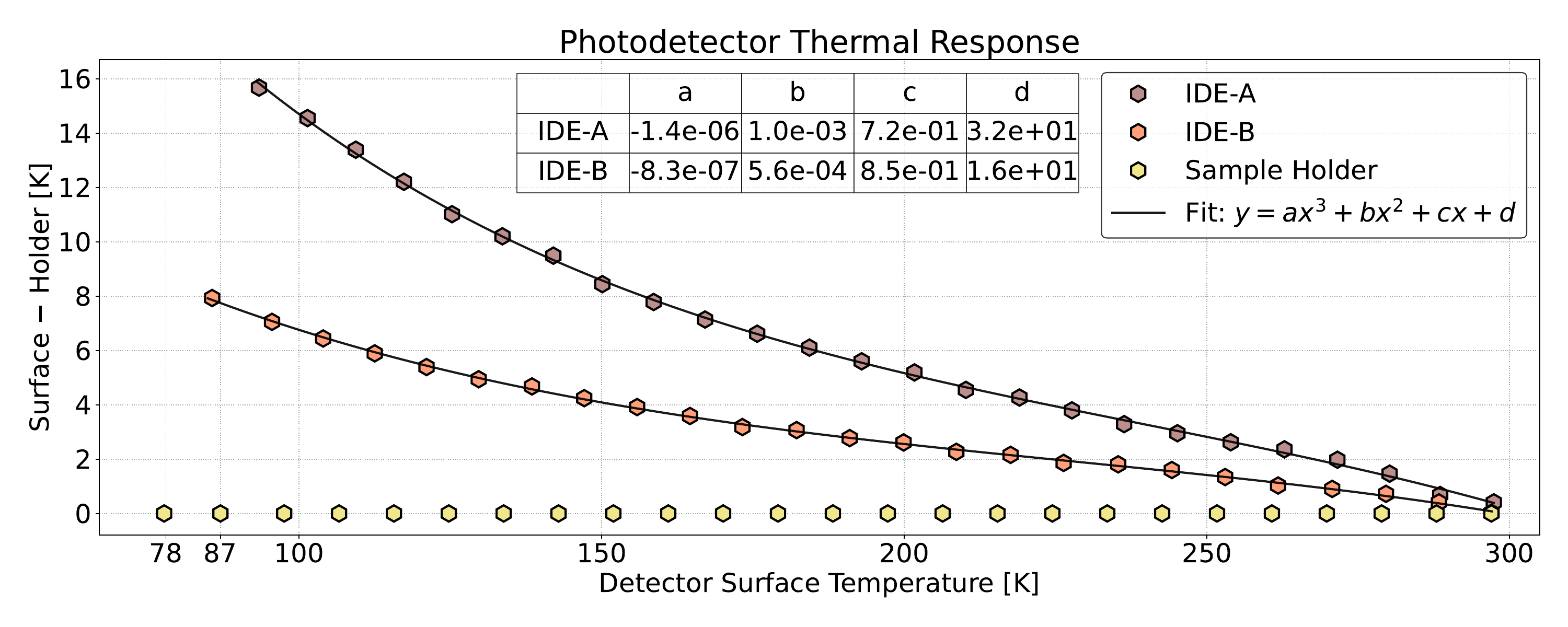}
    \caption{Thermal response of photodetectors in the cryogenic setup. Temperature difference between the detector surface and the sample holder plotted against detector surface temperature for IDE-A and IDE-B with polynomial fits.}

    \label{fig:temps}
\end{figure}

Figure~\ref{fig:temps} shows the temperature difference between the detector surface and the sample holder plotted against the detector surface temperature for both devices. The data were fit with a third-order polynomial of the form $T_{\mathrm{det}} = ax^3 + bx^2 + cx + d$ where \(T_{\mathrm{det}}\) is the detector surface temperature and \(x\) is the holder temperature. Both curves deviate from ideal thermal contact at lower temperatures, with the detector remaining consistently warmer than the holder by several kelvin, indicating measurable thermal resistance between the detectors and the sample holder. IDE-B tracked more closely to the sample holder throughout cooling and reached a minimum temperature of 85.60\,K, whereas IDE-A plateaued at 92.91\,K. This difference arises from slightly improved thermal coupling in IDE-B under otherwise identical operating conditions. The minimum temperatures achieved were limited by thermal contact rather than the cryostat capability, which can reach lower temperatures when operated with liquid helium, though all measurements in this work used liquid nitrogen.

\subsection{Laser Excitation and Photon Flux Control}
\label{sec:laser}

An LPG-405 picosecond pulsed laser from Photek was used as the optical photon source \cite{photek}. The emission wavelength was measured to be 400.76 nm ± 1.05 nm using a CCS200 compact spectrometer \cite{thorlabs}. The laser includes a 16-position pulse width (PW) selector with nominal pulse durations ranging from 35 ps to 120 ps, based on the timing diagram in the datasheet~\cite{photek}. These values were not independently verified and were not used in any calculations. Collimated pulse energies were measured using a calibrated PM16-121 power meter, yielding 8.19 ± 0.41 pJ at the lowest setting and 19.90 ± 1.00 pJ at the highest. A 5\% power meter calibration uncertainty dominates the systematic error. In low-photon-flux experiments, the PW selector, used in combination with absorptive neutral density filters, enabled fine control of the number of incident photons per pulse and is discussed in more detail in Section~\ref{sec:SiPMCalibration}.

The laser was triggered at 4 Hz using a Siglent SDG 1032X function generator in all experiments. Synchronization was achieved using the laser’s sync-out line to trigger the oscilloscope, providing stable timing with 3 ps RMS trigger jitter and 2 ps RMS signal jitter, as specified by the LPG-405 datasheet \cite{photek}. All laser power and signal lines were shielded to minimize electrical noise.

\begin{figure}[htbp]
    \centering
    \includegraphics[width=0.48\textwidth]{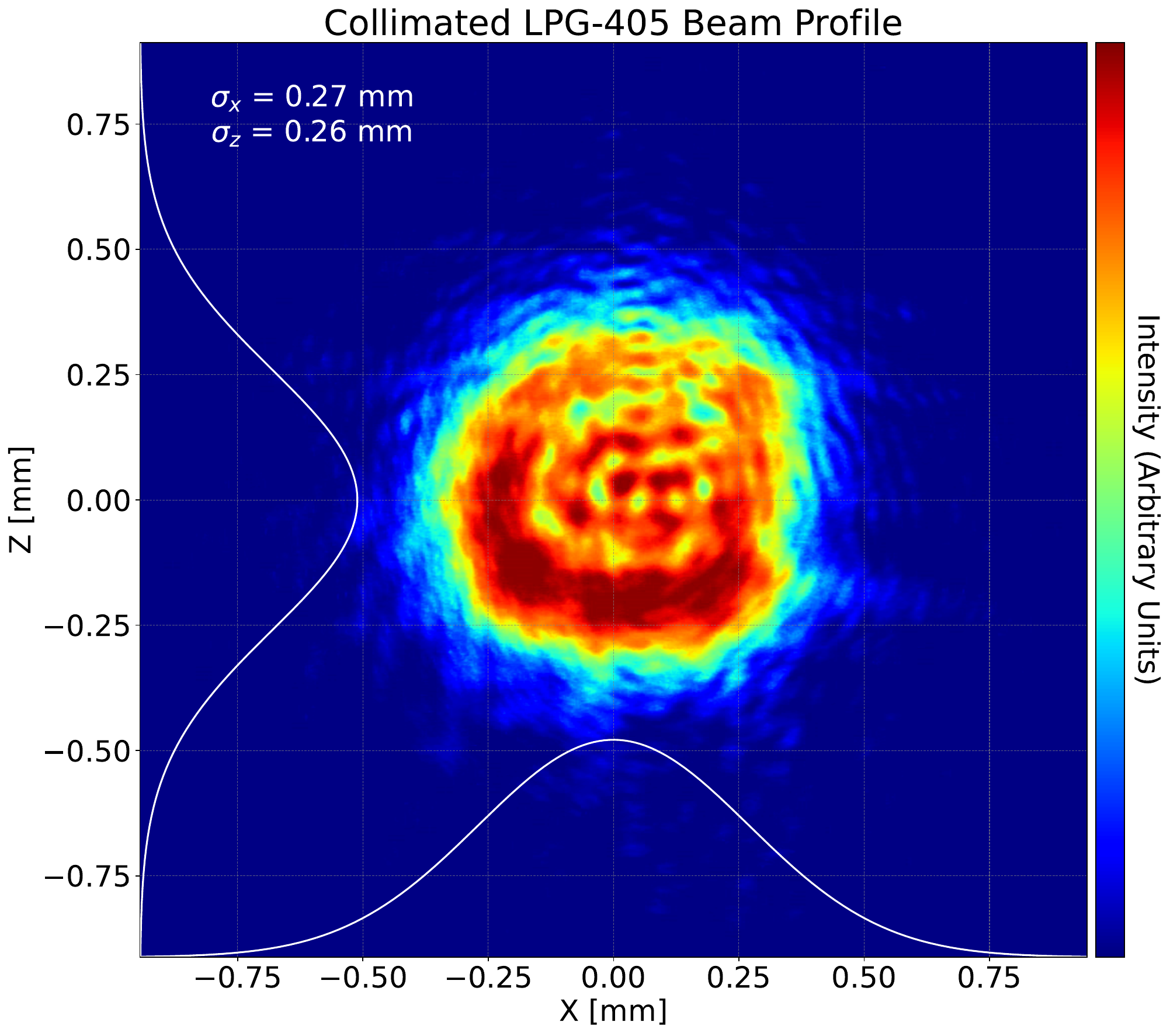}%
    \hspace{3em}%
    \raisebox{2.65ex}{\includegraphics[width=0.377\textwidth]{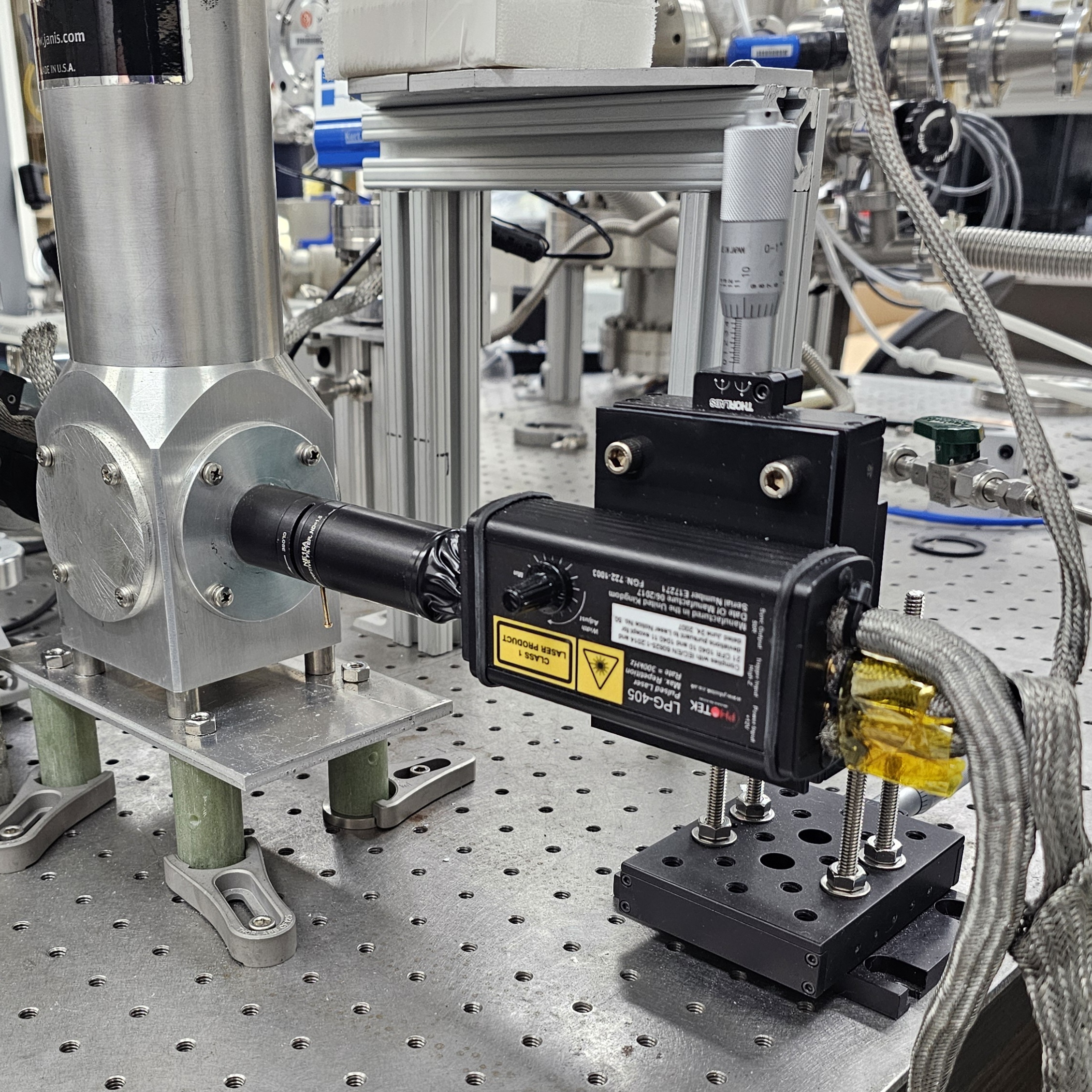}}
    \caption{Left: Heat map of the collimated laser beam profile at the detector plane. The two-dimensional intensity distribution shows a central peak, with standard deviations of 0.27\,mm and 0.26\,mm in the \(x\)- and \(z\)-directions respectively, determined from Gaussian fits to the one-dimensional intensity projections. The full beam diameter at the detector plane is 1.2\,mm. Right: Photograph of the experimental alignment setup showing the laser, micrometer stage, lens tube assembly and optical cryostat interface.}
    \label{fig:beam}
\end{figure}

Beam geometry was defined at the detector plane by adjusting the laser focus to produce a fine circular spot of 3.8\,mm diameter at a working distance of 20\,cm, verified with a laser viewing card. A custom-fabricated acrylic aperture stop with a 1.2\,mm opening was mounted inside the laser barrel to produce a consistent beam profile of 1.2\,mm. The beam was imaged with and without the aperture stop using a Celestron HD digital imager. The protective cover on the imager was removed to eliminate optical distortions, internal reflections, or wavelength filtering. To avoid saturating the imager sensor, a NE09A absorptive neutral density filter was placed in the beam path. The collimated beam profile, shown in Figure~\ref{fig:beam} (left), exhibits a central peak consistent with a Gaussian shape in both x and z projections. The standard deviations of the Gaussian fits to the 1D intensity profiles were 0.27 mm in x and 0.26 mm in z.

The laser assembly was mounted on a pair of precision linear stages providing micrometer-controlled motion in the x- and z-directions. The beamline was aligned to the optical cryostat via a UV-visible viewport fitted with a threaded one-inch lens tube adapter. The complete optical path consisted of a manual shutter, a calibrated neutral density filter stack and additional lens tube segments connecting to the laser barrel. Photon flux at the detector was controlled through a combination of absorptive neutral density filters and adjustment of the laser PW. This dual control allowed fine tuning of incident photon yield over a wide dynamic range without altering the optical alignment. A photograph of the alignment setup is shown in Figure~\ref{fig:beam} (right), where the micrometer stage, lens tube assembly, and optical cryostat interface are visible.

Alignment was performed with the IDE detector installed in the cryostat at operating temperature. The micrometer stages were then adjusted to maximize the photocurrent signal observed on the oscilloscope. Once aligned, the segment of lens tube overlapping the laser barrel was wrapped in Scotch Vinyl Electrical Tape Super 88 to achieve a fully light-tight seal.

\FloatBarrier

\subsection{SiPM Calibration and Photon Counting}
\label{sec:SiPMCalibration}
\begin{figure}[htbp]
    \centering
    \includegraphics[width=0.35\textwidth]{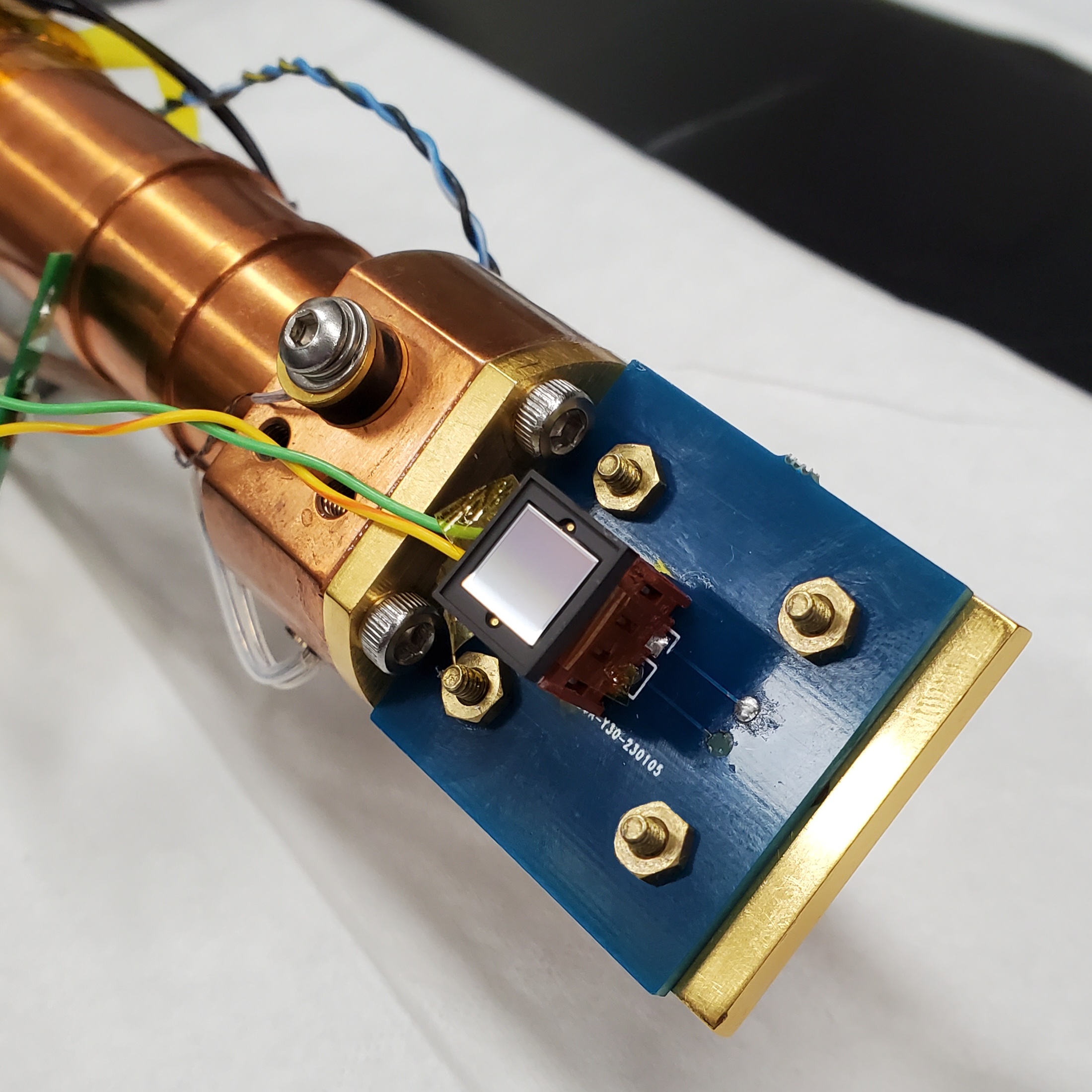}%
    \hspace{1em}%
    \raisebox{2.65ex}{\includegraphics[width=0.58\textwidth]{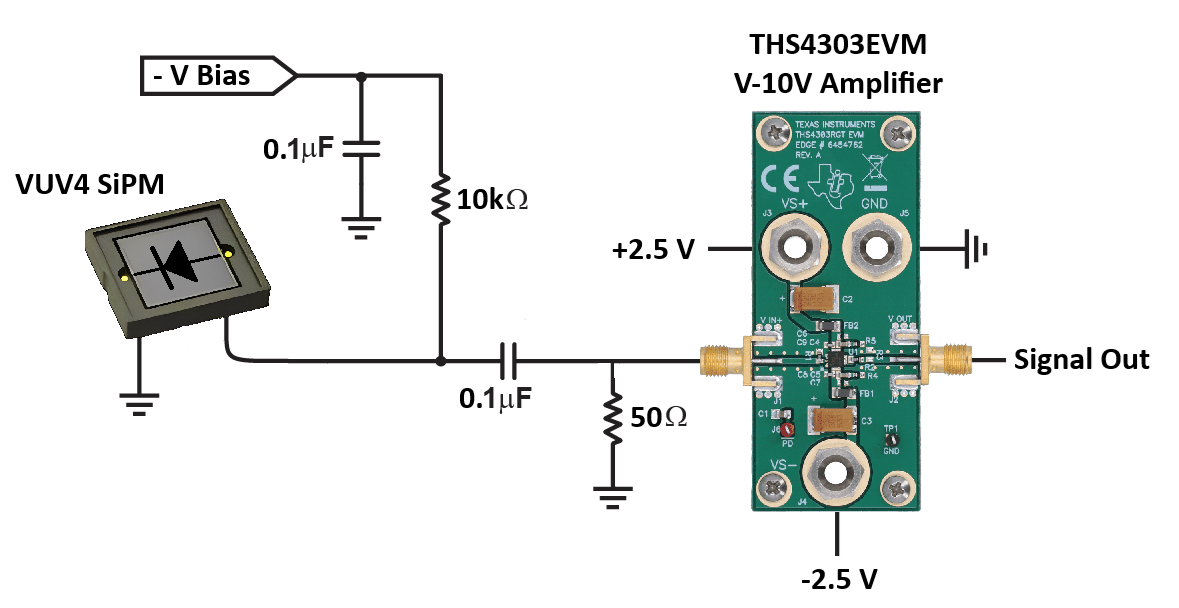}}
    \caption{Photographs of the SiPM calibration setup. Left: The Hamamatsu S13370-6050CN SiPM mounted on the sample holder. Right: The THS4303 amplifier board with an overlaid circuit diagram showing the supporting bias and readout components.}

    \label{fig:sipmpic}
\end{figure}

A Hamamatsu S13370-6050CN VUV4 MPPC was used for all photon flux calibration measurements. The SiPM signal was amplified using a Texas Instruments THS4303 evaluation module (EVM), which incorporates a THS4303 wideband fixed-gain V-10V amplifier. The amplified output was read by a Tektronix 4 Series B MSO oscilloscope. A CAEN DT8301M programmable power supply provided the SiPM bias voltage. Photographs of the mounted SiPM and amplifier circuit are shown in Figure~\ref{fig:sipmpic}. The SiPM has a sensitive area of 6\,$\times$\,6\,mm$^2$, which fully contains the 1.2\,mm diameter laser spot.

Calibration was performed in the same cryostat configuration used for a-Se measurements, with the cold head rotated to place the SiPM in the beam path of the LPG-405 ps laser. To ensure accurate photon flux determination as a function of laser PW and optical filtering, all calibrations were conducted at room temperature. The vacuum chamber was pumped down to the $\upmu$Torr range. Initial laser alignment was performed at a bias of 55 V by pulsing the laser at 10 Hz and adjusting the micrometer stage to maximize the SiPM signal on the oscilloscope.

To resolve single-photon events, the laser flux was minimized by reducing the beam diameter to 1.2 mm and inserting a Thorlabs NE50A-A absorptive neutral density filter. The laser PW was set to its minimum value to yield low detected photon counts per pulse, on the order of single photons.

Calibrations were performed at five bias voltages: 54.0, 54.5, 55.0, 55.5, and 56.0 V. This range enabled determination of the breakdown voltage, V$\text{BR}$ = 51.65 V and validated the datasheet-recommended operating point of V$\text{BR}$ + 4 V = 55.65 V.

\begin{figure}
    \centering
    \includegraphics[width=.8\textwidth]{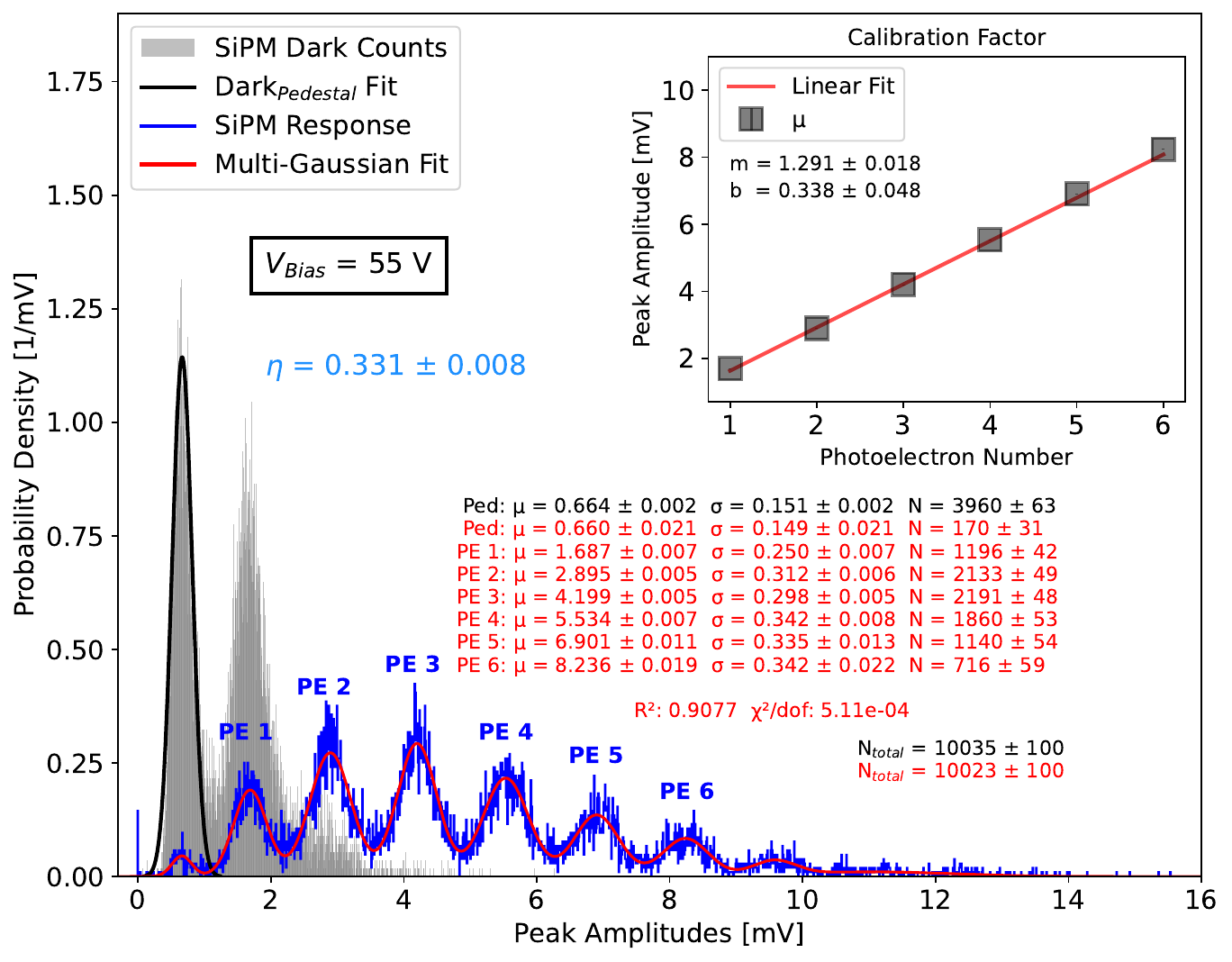}
    \caption{Representative SiPM calibration at 55\,V. Histogrammed peak amplitudes from 10{,}000 dark (gray) and illuminated (blue) waveforms are shown with Gaussian fits to the pedestal and photoelectron peaks (up to 6\,PE). Fit parameters with uncertainties are listed for each peak. The inset displays the linear calibration of peak amplitude versus photoelectron index, with the slope expressed in mV/PE and associated uncertainty. This calibration is used to convert measured SiPM amplitudes to detected photon counts for estimating the incident flux on the a-Se detector. Also shown are the calculated detection efficiency, total event counts and statistical fit quality metrics ($R^2$, $\chi^2/\text{dof}$). The calibration factor is obtained using peak amplitude rather than integrated pulse area; see Section~\ref{sec:SiPMCalibration}.}

    \label{fig:singleCal}
\end{figure}

The laser’s synchronization output was used to trigger the oscilloscope, ensuring stable timing across all measurements. For each bias voltage, 10k dark and 10k illuminated pulses were recorded. Peak amplitudes were extracted using the oscilloscope’s built-in measurement functions and histogrammed. Although SiPM calibrations are conventionally performed using integrated charge, the peak amplitude method was employed here, relying on the repeatability of the pulse shape. This condition was satisfied by the picosecond laser excitation, fixed-gain amplifier, and low noise floor. A comparison with the integrated pulse method confirmed good agreement, as shown in Appendix~\ref{sec:sipmcompare}. The standard charge-based approach is described in~\cite{Eckert2010}. The resulting histograms showed well-defined photoelectron peaks, enabling straightforward extraction of the calibration factor, as illustrated in Figure~\ref{fig:singleCal}. Calibration measurements were repeated periodically and showed minimal variation over time.

To extract SiPM gain, multi-Gaussian fits were applied to resolve discrete photoelectron (PE) peaks. A representative result at 55 V is shown in , including both dark and illuminated histograms, fits to the pedestal and PE peaks (up to 6 PE) and an inset linear fit of PE index versus peak amplitude. The gain, expressed in mV/PE, was taken as the slope of this linear fit. Gain increased monotonically with bias, consistent with expected avalanche behavior. A summary of all calibrations and gain vs. bias trend is included in Figure~\ref{fig:sipmcal} in the appendix.

The average number of photons detected by the SiPM was calculated using the pedestal suppression method: 

\begin{equation}
n = -\ln\left(\frac{N_{\text{ped}}}{N_{\text{tot}}}\right) + \ln\left(\frac{N_{\text{ped,dark}}}{N_{\text{tot,dark}}}\right)
\end{equation}

where \(N_{\text{ped}}\) and \(N_{\text{tot}}\) refer to the number of pedestal events and total events under illumination, respectively, and \(N_{\text{ped,dark}}\) and \(N_{\text{tot,dark}}\) are the corresponding values measured in the dark to correct for thermally induced avalanches. This method estimates the average number of photons detected per pulse by statistically isolating events in which no thermally induced avalanche occurred. Because it is based solely on the pedestal probability, this approach inherently suppresses contributions from optical cross-talk and afterpulsing.

The photon detection efficiency (PDE) $\eta$ was then estimated as 
\begin{equation} \eta = \frac{n}{N_{\text{inc}}} 
\end{equation} 

where N$_{\text{inc}}$ is the estimated number of incident photons. Based on the observation of eight well-resolved photoelectron peaks and a faint ninth, $N_{\text{inc}} \approx 9$ was adopted as the estimated photon count. This yielded a measured PDE of 33.1\% at a bias voltage of 55V for 401 nm illumination. Hamamatsu specifies a nominal peak PDE of 35\% at 500 nm for this device and the lower value at 401 nm is consistent with the expected falloff in sensitivity at shorter wavelengths due to reduced absorption depth and increased surface recombination in silicon~\cite{Piemonte2006}.

The pedestal peak remained stable across all measurements, with mean amplitudes under dark and illuminated conditions of 0.657 ± 0.009 mV and 0.654 ± 0.025 mV, respectively. The corresponding pedestal noise was 0.145 ± 0.009 mV and 0.144 ± 0.018 mV. The intercept of the PE linear fit was consistent across bias values at 0.345 ± 0.020 mV. All multi-Gaussian fits accurately captured the PE peaks, with reduced chi-squared values consistently low at $\chi^2/\text{dof} \sim 5 \times 10^{-4}$, likely reflecting high signal fidelity, low baseline noise, and slightly overestimated fit uncertainties. The coefficient of determination $R^2$ for each PE calibration ranged from 0.9046 to 0.9472. Together, these observations confirm that the PE peaks were well-resolved and that the gain calibration procedure was robust and repeatable.

Several absorptive neutral density filters and combinations, referred to as filter stacks, were used throughout the experiments. The filters included OD40A-A, NE15A, NE09A, NE509B, NE505B and NE503B from Thorlabs. Each stack was coupled to the optical cryostat via a UV-grade fused silica window, which was treated as an additional optical element with independently measured transmission of 0.91500 ± 0.00048.

Transmission measurements for all filters and stacks were performed using the same LPG-405 laser employed in the experiments to ensure consistency. A Thorlabs PM16-121 calibrated power meter was used and all measurements were conducted using the collimated beam configuration. Each filter was characterized individually before being combined into stacks. Transmission measurements were averaged over 100 measurements per filter. While statistical fluctuations were present, the 5\% calibration uncertainty of the power meter dominated the total error budget.

Laser energy stability was characterized over multiple PW settings by measuring the photon flux over 1{,}000 pulses at each setting. Jitter was quantified as the ratio of the standard deviation to the mean ($\sigma/\upmu$) of the measured flux, revealing relative fluctuations decreasing from 4.4\% to 2.2\% as PW increased. This consistent trend indicates that the laser’s energy stability improves at higher output levels and that the dominant source of fluctuation is systematic and intrinsic to the laser rather than detector noise or environmental variation.

A complete listing of filter stacks, total measured transmission and estimated photon flux per pulse is provided in Table~\ref{tab:filter_stacks}. The table is segmented by experiment. The first row corresponds to the highest-transmission stack used in tellurium-doped a-Se measurements. Rows labeled A–I (blue) correspond to stacks used in the response curve experiment and the remaining green-labeled rows were used in the low-photon-flux experiment.

Photon counts for stacks H, I and the four green low-flux stacks were independently measured using the calibrated SiPM, providing a cross-check of the transmission-based photon estimates. The power meter values were consistently higher than those obtained from the SiPM, with an average relative difference of 6.76\%. All values reported include propagated uncertainties and reflect systematic errors dominated by the power meter calibration.

For photon flux levels too high to be accurately measured by the SiPM, the number of incident photons per pulse, $N_{\gamma}$, was estimated using a transmission-based calculation:

\begin{equation}
N_{\gamma} = \left( \prod_i T_i \right) \cdot \frac{E_{\text{pulse}} \cdot \lambda}{hc}
\label{eq:photon_count}
\end{equation}

\noindent where $\prod_i T_i$ is the product of independently measured transmissions for each filter and the UV window, $E_{\text{pulse}}$ is the measured laser pulse energy, $h$ is Planck’s constant, $c$ is the speed of light and $\lambda$ is the laser wavelength. All flux estimates using this method carry a 5\% systematic uncertainty dominated by the calibration error of the PM16-121 power meter.

Calibrated SiPM measurements were used at photon flux levels within the device’s linear response range. For each optical filter setting yielding incident photon flux below approximately 4.5k photons per pulse, a histogram of SiPM pulse amplitudes was acquired and fit with a Gaussian. This linearity limit, within 5\%, is set by the SiPM’s dynamic range of 14.4k microcells~\cite{Hamamatsu2021,HamamatsuVUV4}. The number of incident photons was then calculated using

\begin{equation}
N_{\gamma} = \frac{\mu}{G \cdot \eta}
\pm N_{\gamma} \cdot \sqrt{
\left( \frac{\delta\upmu}{\mu} \right)^2 +
\left( \frac{\delta G}{G} \right)^2 +
\left( \frac{\delta \eta}{\eta} \right)^2
}
\label{eq:photon_conversion}
\end{equation}

\noindent Here, $N_{\gamma}$ is the estimated number of photons per pulse, $\upmu$ is the Gaussian mean of the pulse amplitude distribution, $G$ is the calibration factor in mV per photoelectron and $\eta$ is the photon detection efficiency. The uncertainty in $N_{\gamma}$ is propagated from the fractional uncertainties in each input parameter, derived from the calibration fits described previously.

\begin{table}[htb]
\centering
\begin{tabular}{|c|c|c|c|c|}
\hline
\textbf{ } & \textbf{Filter Stack} & \textbf{PW} & $\boldsymbol{\prod_i T_i}$ & $\boldsymbol{N_\gamma}$ \\
\hline
\rowcolor{red!10}
 $\alpha$ & NE09A & min & \texttt{9.75e-2} & 1,565,763 \\
\hline
\rowcolor{cyan!5}
A & NE15A & max & \texttt{2.31e-2} & 954,683 \\
\hline
\rowcolor{cyan!5}
B & NE15A + NE503B & max & \texttt{1.44e-2} & 547,420 \\
\hline
\rowcolor{cyan!5}
C & NE15A & min & \texttt{2.31e-2} & 370,668 \\
\hline
\rowcolor{cyan!5}
D & NE15A + NE503B & min & \texttt{1.32e-2} & 211,978 \\
\hline
\rowcolor{cyan!5}
E & NE15A + NE505B & min & \texttt{5.99e-3} & 96,259 \\
\hline
\rowcolor{cyan!5}
F & NE15A + NE09A + NE503B & min & \texttt{1.41e-3} & 22,599 \\
\hline

\rowcolor{cyan!5}
G & NE15A + NE09A + NE509B + NE503B & max & \texttt{1.42e-4} & 5,888 \\
\hline
\rowcolor{cyan!5}
H & NE15A + NE09A + NE509B + NE505B & max & \texttt{6.43e-5} & 2,666\textsuperscript{\dag} \\
\hline
\rowcolor{cyan!5}
I & NE15A + NE09A + NE509B + NE505B & min & 6.43e-5 & 1,034\textsuperscript{\dag} \\
\hline
\rowcolor{green!5}
 L1 & NE15A + NE09A + NE509B & max & 2.47e-4 & 10,405 \\
\hline
\rowcolor{green!5}
 L2 & NE15A + NE09A + NE509B & min & 1.42e-4 & 4,003 \\
\hline
\rowcolor{green!5}
 L3 & NE15A + NE09A + NE509B + NE505B + NE503B & max & 3.68e-5 & 1,529\textsuperscript{\dag} \\
\hline
\rowcolor{green!5}
 L4 & NE15A + NE09A + NE509B + NE505B + NE503B & min & 3.68e-5 & 591\textsuperscript{\dag} \\
\hline
\rowcolor{green!5}
L5 & OD40A-A & min & 1.14e-5 & 183\textsuperscript{\dag} \\
\hline
\rowcolor{green!5}
L6 & OD40A-A & max & 1.14e-5 & 473\textsuperscript{\dag} \\
\hline
\end{tabular}
\caption{Filter stacks used in the Tellurium-doped a-Se response comparison (red), high-field response (blue), and low-photon experiments (green). Transmission values, $\prod_i T_i$ include a 91.5\% contribution from the UV fused silica window. PW selector settings correspond to nominal pulse widths ranging approximately from 35~ps (min) to 130~ps (max), based on the datasheet timing diagram~\cite{photek}. The final column gives the incident photons per pulse, $N_\gamma$. \textsuperscript{\dag}Indicates photon counts derived from SiPM measurements; all others were calculated using Equation~\ref{eq:photon_count} based on power meter measurements. All photon counts include a 5\% uncertainty.}
\label{tab:filter_stacks}
\end{table}

\subsection{Readout Electronics and Signal Acquisition}
\label{sec:readout}

Detector mounting and cryostat integration are described in Section~\ref{sec:cryostat}. Due to the high bias voltages used, up to 3\,kV, the electrical connection to the detector was made using an SHV-5 feedthrough mounted on the optical cryostat. The detector was biased and read out using a CR-110-R2.2 charge-sensitive preamplifier (CSP) installed on a CR-150-R5 evaluation board housed inside a sealed aluminum enclosure.

\begin{figure}[htbp]
    \centering
    \includegraphics[width=0.35\textwidth, trim=150 190 150 150, clip]{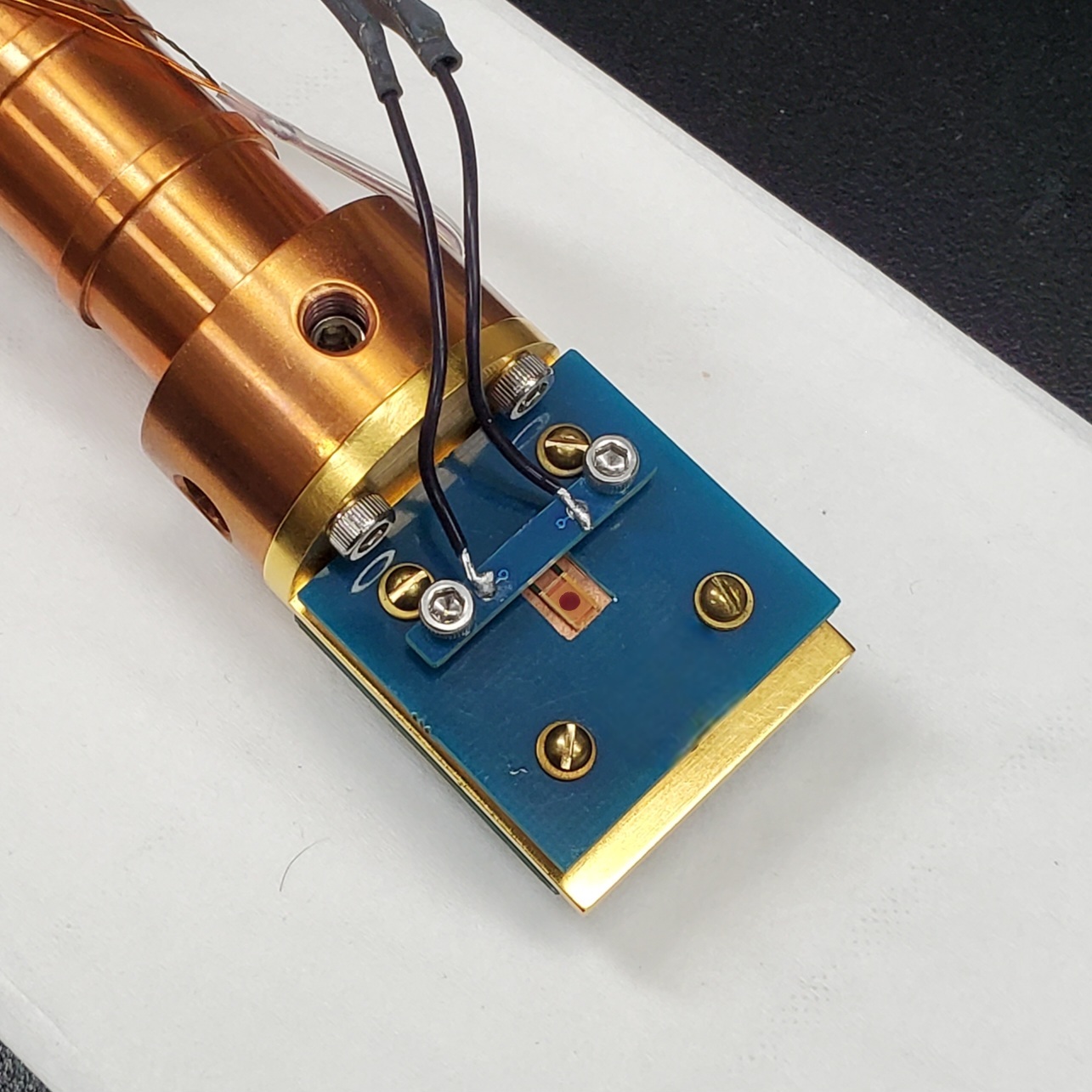}%
    \hspace{1em}%
    \raisebox{2.65ex}{\includegraphics[width=0.58\textwidth]{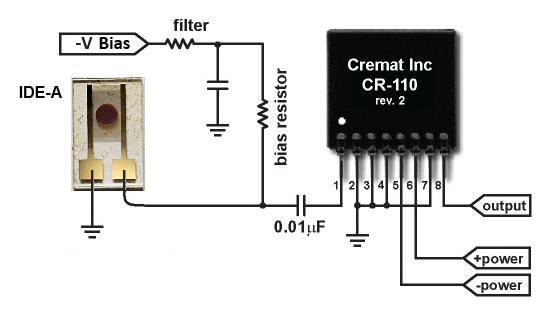}}
    \caption{Photographs of the a-Se setup. Left: An IDE-A sample mounted in its holder. Right: The CR-110 charge-sensitive preamplifier with an overlaid circuit diagram showing the supporting bias and readout components adapted from~\cite{cremat2025}.}

    \label{fig:aSepic}
\end{figure}

To prevent high-voltage discharge, several modifications were made to the CSP housing. An unused copper trace near the high-voltage connection was removed, solder joints were filed to eliminate sharp edges and the interior of the housing was lined with Kapton tape. These mitigations were developed during troubleshooting of intermittent discharge events and are noted here to support reproducibility. Signal output from the CSP was routed to a female SHV-5 jack mounted on the enclosure, allowing direct coupling to the cryostat feedthrough and eliminating the need for intermediate cabling. The high-voltage bias line was connected through a second SHV-5 jack. A photograph of an IDE-A detector installed in the sample holder and a schematic of the CSP connection and decoupling circuit are shown in Figure~\ref{fig:aSepic}.

Noise and grounding were critical considerations. The detector was highly sensitive to mechanical vibration, especially under high bias. Within the cryostat, the high-voltage cable was secured in place while avoiding direct contact with the cold finger to minimize vibrational coupling during thermal contraction. Biasing was only applied after the cryostat reached thermal equilibrium to prevent transient spikes caused by settling of internal components.

All instrumentation and the optical cryostat shared a common ground. Cables connected to the LPG-405 laser were wrapped in ground braid to suppress radiated noise. Signals from the CSP were sent to a Tektronix 4 Series B MSO mixed signal oscilloscope via BNC using a 1\,M$\Omega$ AC-coupled input. The oscilloscope was triggered on a separate channel using the synchronization output from the LPG-405 laser. Both input channels used a 20\,MHz analog bandwidth limit. Given the 140\,$\upmu$s decay time of the CR-110, the horizontal acquisition window was set to 200\,$\upmu$s. Vertical scale was adjusted to accommodate the expected pulse amplitude. Waveforms were sampled at 625\,MS/s and transferred to a connected laptop over Ethernet upon triggering.

The noise performance of the readout chain was characterized to quantify its contribution to the waveform baseline. This analysis was performed using waveform acquisitions collected at 297~K, 165~K, 93~K, and 87~K, as described in Section~\ref{sec:datataking}, to evaluate temperature-dependent behavior across the full operating range. Baseline noise was measured from the pre-signal region of each acquisition window and converted to equivalent noise charge (ENC) based on the CR-110 gain of 1400 $\mathrm{mV/pC}$. Across all fields, the RMS noise at 297~K remained approximately constant, with a mean ENC of $2804 \pm 226~\mathrm{e^{-}}$. A gradual reduction in ENC was observed with decreasing temperature, with mean values of $2509 \pm 170~\mathrm{e^{-}}$ at 165~K, $2466 \pm 306~\mathrm{e^{-}}$ at 93~K, and $2264 \pm 134~\mathrm{e^{-}}$ at 87~K. These values reflect consistent noise performance across field and temperature, with a modest reduction at lower temperatures. The total input capacitance was minimized through direct SHV coupling between the preamplifier and the cryostat feedthrough. The detector capacitance was estimated to be 0.2–0.6 pF, with the remaining contribution dominated by cabling and SHV feedthrough capacitance, totaling approximately 20 pF. The measured ENC exceeds the CR-110 preamplifier’s intrinsic specification of $200~\mathrm{e^{-}}$ RMS, primarily due to contributions from the evaluation board and the amplifier’s sensitivity to environmental perturbations. 

\subsection{Pulsed VUV Argon Flashlamp Configuration}
\label{sec:ArLamp}

A custom argon flashlamp was used as a source of pulsed VUV light~\cite{McDonald2022}. The lamp produces broadband emission from neutral excited argon in the 100--200\,nm range via spontaneous deexcitation in a low-pressure gas discharge. The dominant spectral lines appear at 117.7\,nm, 126.7\,nm, and 129.8\,nm, with additional weaker emission observed near 153\,nm~\cite{McDonald2022}. Although the 128\,nm excimer line associated with liquid argon scintillation is absent, nearby wavelengths can be isolated to probe direct VUV sensitivity of the a-Se detector.

\begin{figure}[htb]
\centering
\scalebox{0.95}{
\begin{tikzpicture}

  \node[inner sep=0pt] at (-5, 0)
    {\includegraphics[width=0.25\textwidth]{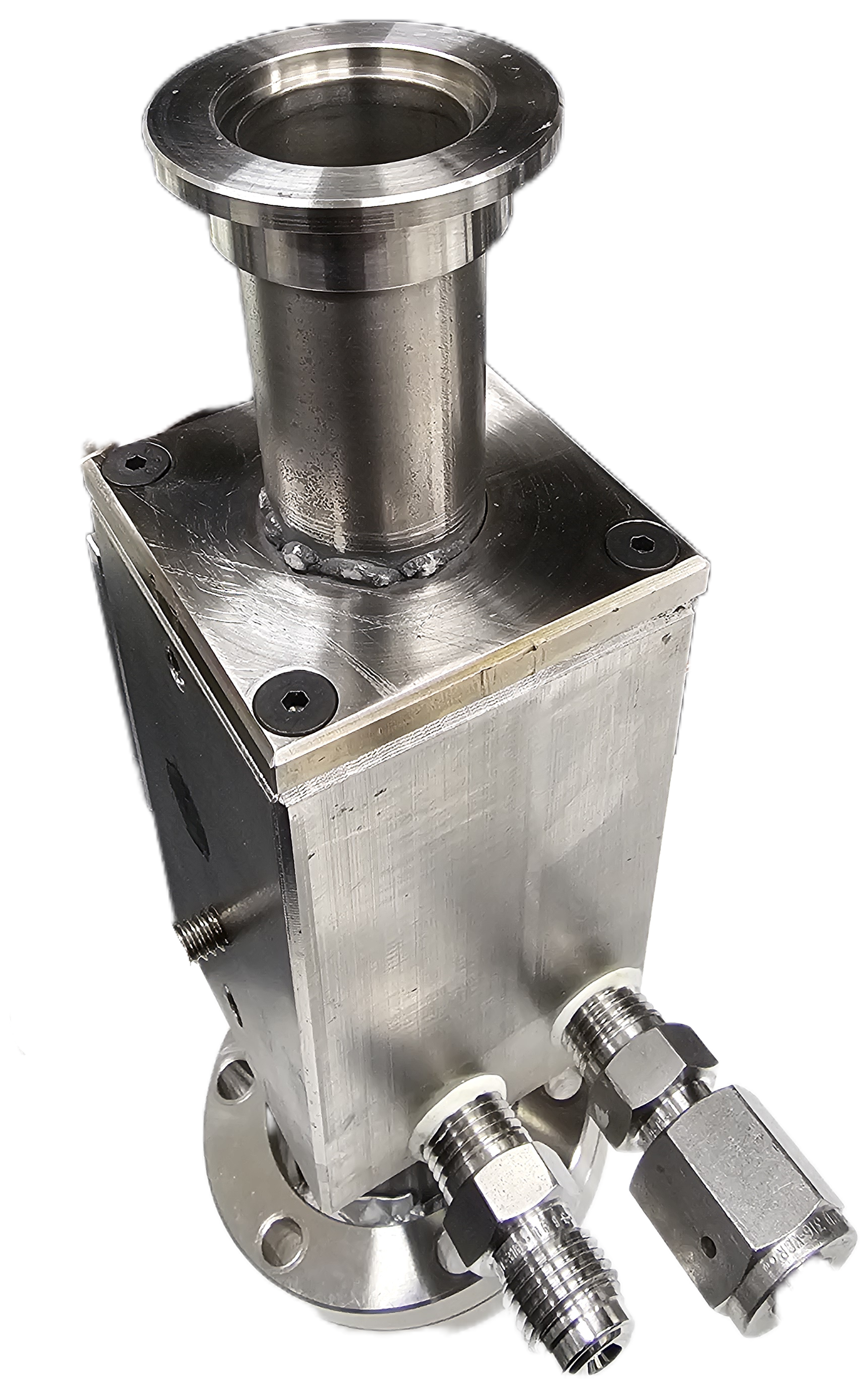}};

  \node[inner sep=0pt] at (0, 0)
    {\includegraphics[width=0.25\textwidth]{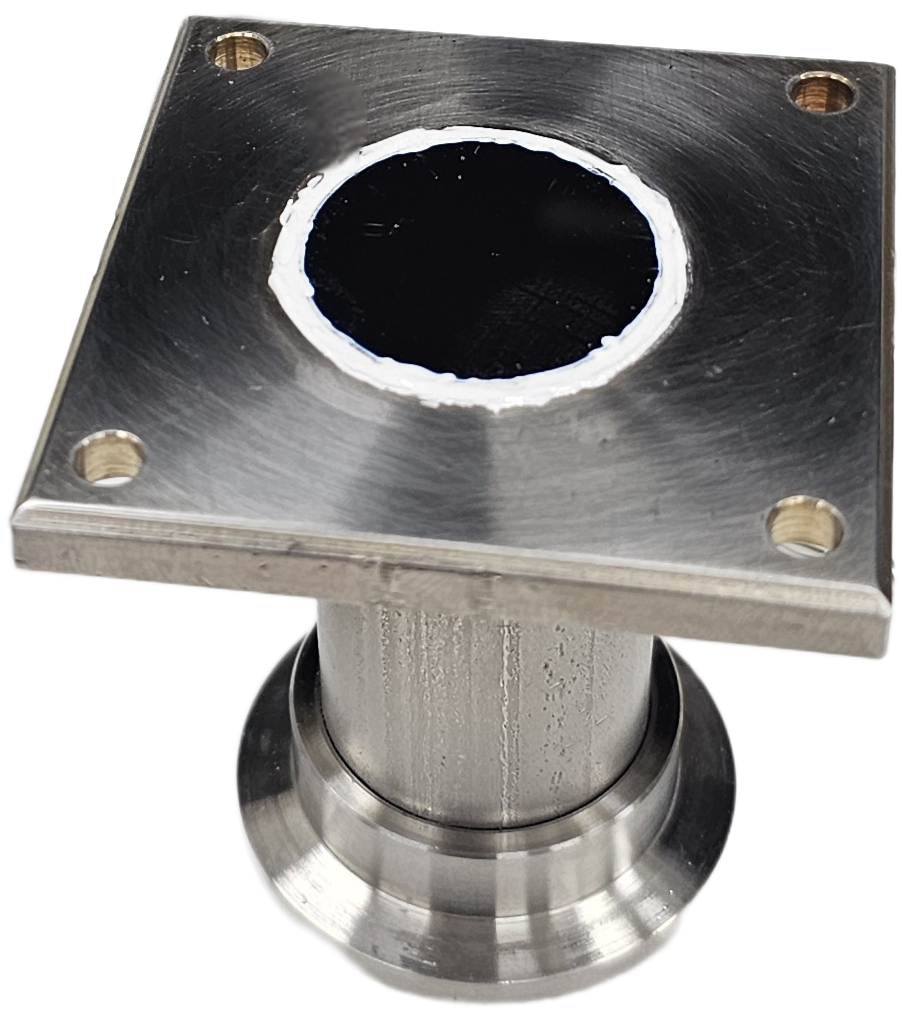}};

  \node[inner sep=0pt] at (5, 0)
    {\includegraphics[width=0.15\textwidth]{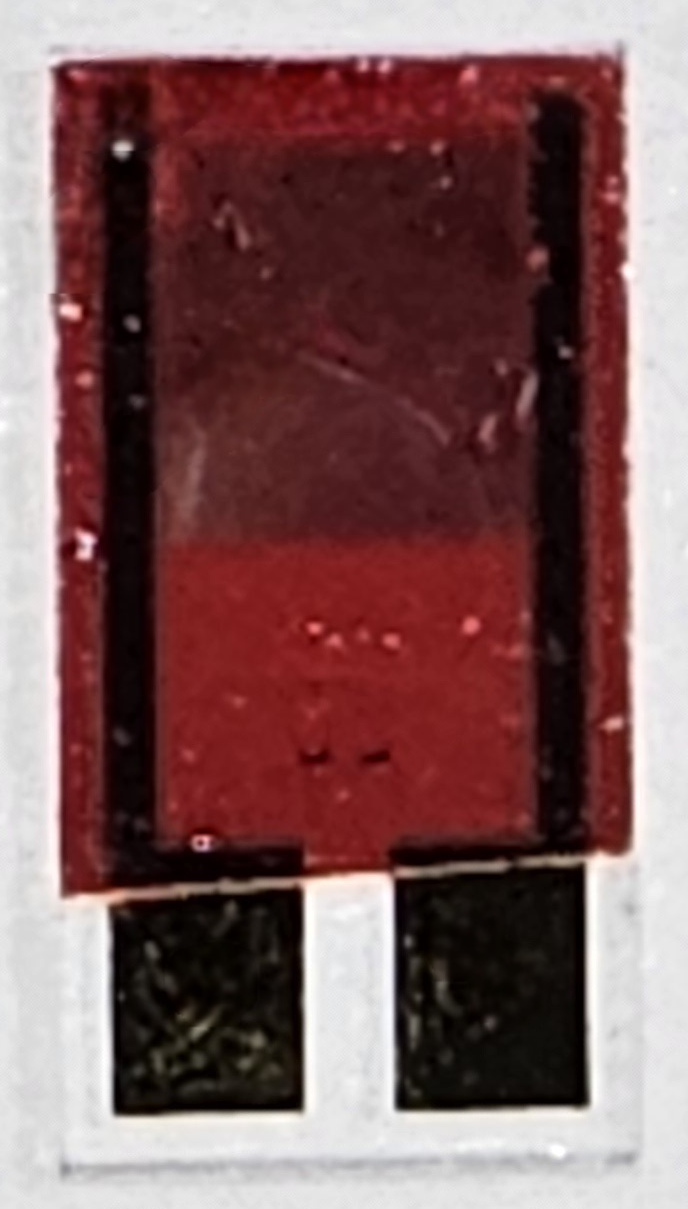}};
  \node at (5, -2.7) {IDE-B};

\end{tikzpicture}
}
\caption{Modified argon flashlamp system and detector used with the setup. 
Left: Spark chamber with the KF25 extension mounted on the axial port. 
Center: Custom filter housing containing the 130\,nm narrowband window.
Right: IDE-B detector fully coated with a-Se.}

\label{fig:Arlamp}
\end{figure}

To couple the lamp output to the detector, the spark chamber was modified with an axial output port aligned on the discharge axis. A KF25 half nipple was welded to the center of an unused axial viewport and a 25130FNB VUV narrowband filter from eSource Optics was epoxied into the end of the 1-inch diameter tube at the spark chamber interface~\cite{esourceoptics2025}. The filter has a peak wavelength of 130\,nm~$\pm$~2.5\,nm, a full width at half maximum of 20\,nm~$\pm$~5\,nm and a peak transmission of 15\%. Both the 126.7\,nm and 129.8\,nm argon emission lines fall within the filter passband, with the 129.8\,nm line near the peak transmission and the 126.7\,nm line transmitted at slightly reduced efficiency. Contributions from weaker spectral lines outside the FWHM, such as those near 117.7\,nm and 153\,nm, are further suppressed but not entirely eliminated. The filter also serves as a physical vacuum barrier, isolating the spark chamber from the external beamline while maintaining spectral purity by suppressing out-of-band transmission above 200\,nm to below 0.1\%. Figure~\ref{fig:Arlamp} shows the modified lamp spark chamber, with the axial output port and KF25 extension facing upward in the assembled view (left), and the custom filter housing with the 130\,nm window visible (right).

The lamp was mounted to a six-way KF25 cross that connected directly to a blank viewport flange on the optical cryostat. This viewport does not contain a window and is maintained under high vacuum, allowing direct VUV transmission to the detector. The six-way cross also provided ports for high-vacuum pumping and gas handling. A KJLC 300 Series convection-enhanced Pirani gauge monitored system pressure. Gas lines were connected via a 1/4-inch compression fitting and T-valve, allowing the spark chamber to be pumped out and isolated from the beamline. After reaching pressures in the $\upmu$\text{Torr}
 range, the vacuum valve was closed and the chamber was backfilled with approximately 5\,Torr of research-grade argon from Airgas.

The flashlamp was operated in driver mode with a discharge rate of 1\,Hz. A bias voltage of approximately +800\,V was applied across the lamp electrodes to initiate the discharge. The resulting light pulse had a duration of approximately 4\,$\upmu$s. The cold head was rotated 180 degrees from the laser configuration so that the detector faced the flashlamp directly. Signals were read out using the same charge-sensitive preamplifier and oscilloscope setup described in Section~\ref{sec:readout}. The oscilloscope was triggered directly on the signal channel using a threshold set just above the noise level. No flux calibration of the flashlamp was performed, as the measurement focused solely on establishing detector response to 130\,nm VUV illumination rather than extracting absolute detection efficiency.

The detector used for this measurement was IDE-B. This detector used the same polyimide blocking layer and followed the fabrication procedure described in Section~\ref{sec:fabrication}, but the a-Se layer was deposited over the entire finger region. This geometry increased the photosensitive area and reduced the likelihood of direct photoelectric response from uncoated gold electrodes.

\subsection{Data Taking Procedure}\label{sec:datataking}
Throughout this section, the term “applied field” refers to the lateral electric field calculated as the applied bias divided by the inter-electrode gap. All measurements were conducted in high vacuum using the optical cryostat and sample holder described previously. For each experiment, the detector was installed and the cryostat was pumped down to the $\upmu$Torr range. Laser-based experiments followed a standard alignment procedure performed at room temperature using the NE09A filter stack labeled $\alpha$ in Table~\ref{tab:filter_stacks}. The detector was biased to $-200$\,V at 1\,V/s while the laser was pulsed at 4\,Hz, and the x–z translation micrometers were adjusted to maximize the signal amplitude on the oscilloscope. Following alignment, the bias was returned to zero at the same ramp rate. 

Temperature was adjusted by slowly introducing liquid nitrogen into the ST-100 cold head until the desired set point was reached. Once the temperature stabilized to within 10\,mK, any residual surface charge was cleared by momentarily applying a small positive bias of $+10$\,V, followed by a return to zero before ramping to the target negative bias. All voltage ramps were performed at 1\,V/s up or down.

\paragraph*{Low-Photon Threshold Experiment}

This measurement used the IDE-B detector. Two temperatures were investigated: 87\,K and 165\,K. Bias voltages of 2600\,V and 2800\,V were applied corresponding to applied fields of 110.7\,V/\textmu m and 119.2\,V/\textmu m, respectively.

At each bias, the detector was held in the dark for one minute prior to illumination. The function generator was configured to trigger the laser 100 times at 4\,Hz. Beginning with the minimum PW setting, the shutter was opened and waveforms were recorded on the oscilloscope. The PW selector was then incremented and data acquired for all 16 available settings, corresponding to nominal pulse widths from 35 to 120 ps as described in Section~\ref{sec:laser}. For selected bias voltages, the filter stack was exchanged to reduce photon flux and the measurement procedure repeated. After each run, the detector bias was ramped to zero and the system allowed to return to room temperature under vacuum before repeating the procedure at a different temperature.

\paragraph*{High-Field Response Curve Experiment}

The IDE-A detector was used for this study. The measurement was performed at 93\,K and repeated two days later using the same detector to assess reproducibility. Both runs followed an identical procedure using the PW-stack combinations labeled A through I in Table~\ref{tab:filter_stacks}, which span a wide dynamic range in photon flux. The bias was varied from 1400\,V to 2400\,V in 200\,V steps, corresponding to applied fields from 68.1\,V/\textmu m to 116.7\,V/\textmu m.

For each bias setting, the detector was held at field in the dark for one minute. The laser was pulsed 50–100 times at 4\,Hz and waveforms recorded for each of the A–I filter combinations. Stacks were changed sequentially from high to low transmission. For the first run, this procedure was performed up to 107.0\,V/\textmu m. For the second run, data was collected at the same bias points and an additional field of 116.7\,V/\textmu m.

Between runs, the cryostat was brought to atmospheric pressure at room temperature, the detector was removed and reinstalled and the system was re-pumped and re-aligned. The detector then sat in vacuum for 2 days before the repeat measurement.

\paragraph*{VUV Flashlamp Illumination}

This measurement used the fully-coated IDE-B detector shown in Figure~\ref{fig:Arlamp}. Measurements were performed at 297\,K and 87\,K. The detector was installed, the system pumped down and held in the dark for 12 hours prior to cooling. The cold head was rotated 180 degrees to align the detector with the flashlamp beamline.

Bias voltages ranged from 900\,V to 2800\,V, corresponding to applied fields from 38.3\,V/\textmu m to 119.2\,V/\textmu m. For the room temperature run, the detector was biased to 900\,V and held for one minute before triggering the flashlamp at 1\,Hz. Waveforms were recorded on the oscilloscope using a threshold trigger slightly above noise level. The bias was increased in 100\,V steps and the procedure repeated up to 1900\,V.

After the room temperature run, the detector was removed and reinstalled. The system was evacuated, held in the dark for 12 hours, cooled to 87\,K and the measurement repeated up to 2800\,V. In both runs, approximately 20 waveforms were recorded per bias point.

\paragraph*{Tellurium-Doped a-Se Comparison}

This measurement used three IDE-B detectors fabricated with a-SeTe and two IDE-A detectors fabricated with undoped a-Se. Two temperatures were selected: 297 K and 93 K. While IDE-B devices are capable of reaching 87 K, the 93 K point was chosen to enable direct comparison with undoped a-Se data previously collected using IDE-A devices. In that earlier work, bias voltages were applied in 100 V steps. For the present measurements, bias voltages were selected to reproduce the corresponding applied field values from the undoped dataset.

Each detector followed the same preparation as in previous experiments. For both the 297\,K and 93\,K data sets, the bias was increased in steps chosen to match the prior applied fields, reaching a maximum of 82.7\,V/\textmu m at 297\,K and 102.1\,V/\textmu m at 93\,K. At each setting, the detector was biased, held for one minute, and the laser triggered to record 90 waveforms. Between each measurement, the bias was brought to zero and surface charge cleared with a +10\,V pulse. A new detector was used for each run, with two runs performed at 297\,K and one at 93\,K.

\section{Results and Discussion}\label{sec:results}

\begin{table}[htb]
\centering
\small
\renewcommand{\arraystretch}{1.4}
\begin{tabular}{@{}l l l l p{4.6cm}@{}}
\toprule
\textbf{Experiment} & \textbf{Detector} & \makecell[c]{\textbf{Temperature} \\ \textbf{[K]}} & \makecell[c]{\textbf{Field Range} \\ \textbf{[V/\(\upmu\)m]}} & \textbf{Experimental Objective} \\

\midrule
\makecell[l]{Low-Photon \\ Threshold} & IDE-B & 87\,K, 165\,K & 110.7–119.2 & Determine single-shot detection threshold using matched filter analysis \\

\makecell[l]{High-Field \\ Response} & IDE-A & 93\,K & 68.1–116.7 & Assess linearity and onset of avalanche gain across wide flux range \\

\makecell[l]{VUV \\ Flashlamp} & \makecell[l]{IDE-B \\ (fully coated)} & 87\,K, 297\,K & 38.3–119.2 & Evaluate direct response to narrowband 130\,nm VUV light \\

\makecell[l]{a-SeTe \\ Comparison} & \makecell[l]{IDE-A a-Se \\ IDE-B a-SeTe} & 93\,K, 297\,K & \makecell[l]{4.8-102.1} & Compare avalanche and transport behavior of doped vs.\ undoped a-Se \\
\bottomrule
\end{tabular}
\caption{Summary of the four experiments, showing detector type, operating temperature, range of applied electric fields and experimental objectives.}
\label{tab:summary_experiments}
\end{table}

This section presents results from the full set of experiments, summarized in Table~\ref{tab:summary_experiments}. It begins with a description of the waveform-fitting methodology used throughout the analysis, followed by a geometric evaluation of the photosensitive area and the definitions of quantitative efficiency metrics. Section~\ref{sec:LowP} characterizes detector performance in the low-photon regime at 401 nm to establish practical sensitivity limits. The response linearity and onset of avalanche multiplication at 93 K are examined in Section~\ref{sec:linearity_response}. Section~\ref{sec:direct_vuv} reports measurements demonstrating direct detection of narrow-band 130 nm VUV light at both room and cryogenic temperatures. Section~\ref{sec:Te_aSe_comparison} presents results from detectors fabricated with tellurium-doped a-Se and compares them to measurements using undoped a-Se, highlighting material-dependent differences in charge transport and gain under cryogenic conditions.

\subsection*{Waveform Modeling and Fit Procedure}
\label{sec:WaveformModel}

To extract quantitative features from individual detector waveforms, each signal was fit using a model designed to capture the convolved response of the a-Se photodetector and the CR-110 charge-sensitive preamplifier. The functional form is
\begin{equation}
f(t) = y_0 + A \left(1 - e^{-(t - x_0)/t_1}\right)^p \Theta(t - x_0) e^{-(t - x_0)/t_2}
\end{equation}
where \( \Theta(t - x_0) \) is the Heaviside step function, enforcing causality by setting the signal to zero for \( t < x_0 \). The parameters \( y_0 \), \( A \) and \( x_0 \) represent the baseline offset, peak amplitude and signal onset time, respectively. The parameters \( t_1 \) and \( t_2 \) describe the characteristic rise and decay times and the shaping exponent \( p \) adjusts the sharpness of the peak. Representative waveforms and their fits using this model are shown in Figures~\ref{fig:matched1} and~\ref{fig:matched2}.

Prior to fitting, waveforms were baseline-subtracted using a pre-trigger region average and filtered in the frequency domain with a low-pass cutoff at 150\,kHz to suppress high-frequency noise. The time window was typically restricted to \(-15~\upmu\text{s} \leq t \leq 150~\upmu\text{s}\) to isolate the relevant pulse features for a small signal though the domain was dynamically extended for larger signals to ensure full capture of the waveform shape.

Fitting was performed in two stages. A global optimization using differential evolution was first employed to explore the parameter space and avoid local minima. The result was then refined by local minimization using the limited-memory Broyden–Fletcher–Goldfarb–Shanno algorithm with bound constraints (L-BFGS-B) \cite{byrd1995}. To improve robustness against noise and outliers in low-amplitude waveforms, the objective function minimized a Huber loss~\cite{huber1964}.

Fits were subjected to quality checks and rejected if either of the following conditions was met. First, the fitted waveform had zero amplitude, indicating failure to resolve a signal. Second, the fitted baseline region ($-15~\upmu\text{s} < t < -10~\upmu\text{s}$) exhibited a slope magnitude exceeding $10^{-3}~\text{mV}/\upmu\text{s}$, indicating distortion or poor convergence near the pre-pulse baseline.

For accepted fits, the following waveform properties were extracted: the peak amplitude relative to the baseline, rise time between 10\% and 90\% of the peak height and signal-to-noise ratio (SNR), computed as the ratio of the squared fitted signal to the squared residuals between the data and the fit. These quantities, along with the best-fit parameters, were recorded for subsequent analysis.

\subsection*{Active Area and Geometric Fill Factor}
\label{sec:fillFactor}

In laterally biased photodetectors with IDEs, photogenerated holes only contribute to the measured signal if they originate in regions with sufficient electric field to enable drift. In these devices, the electric field is primarily confined to the gaps between opposing metal fingers. Regions of a-Se located directly above the electrode fingers are shielded from the field and do not contribute appreciably to carrier collection.

To first order, the spatial sensitivity of the detector can be described by the geometric fill factor \(\phi_g\), defined as the ratio of the gap width \(w_{\text{gap}}\) to the total pitch of a single electrode pair, given by the sum of the gap and finger widths \(w_{\text{gap}} + w_{\text{finger}}\):
\begin{equation}
\phi_g = \frac{w_{\text{gap}}}{w_{\text{gap}} + w_{\text{finger}}}.
\end{equation}

This expression is purely geometric and does not account for any field fringing or voltage-dependent effects, which are expected to be small compared to the fill factor and are neglected here. When the a-Se sensing region is uniformly illuminated, the number of photons incident on field-accessible regions is approximately
\begin{equation}
N_{\text{field}} = \phi_g \cdot N_{\text{ph}},
\end{equation}
where $N_{\text{ph}}$ is the total number of incident photons.

In the measurements described here, the detectors were illuminated with a two-dimensional Gaussian laser beam centered on a circular dot of a-Se. The intensity distribution of the beam is given by
\begin{equation}
I(x,z) \propto \exp\left(-\frac{x^2}{2\sigma_x^2} - \frac{z^2}{2\sigma_z^2}\right),
\end{equation}
where $\sigma_x$ and $\sigma_z$ describe the spatial spread of the beam in the $x$ and $z$ directions, respectively. While the interdigitated field structure is uniform near the center of the dot, it becomes truncated near the edges. To account for both the spatial variation in illumination and the nonuniformity in the local fill factor, the effective fill factor can be defined as:
\begin{equation}
\phi_{\text{eff}} = \frac{\iint_{A_{\text{dot}}} I(x,z)\,f(x,z)\,\mathrm{d}x\,\mathrm{d}z}{\iint_{A_{\text{dot}}} I(x,z)\,\mathrm{d}x\,\mathrm{d}z},
\end{equation}
where $f(x,z)$ is the local geometric fill factor. In the central region, $f(x,z) = \phi_g$, while near the edges it tapers to zero due to partial truncation of the electrode pattern. However, the Gaussian weighting suppresses the contribution of the edge region by more than an order of magnitude, and numerical evaluation of the integral shows that $\phi_{\text{eff}}$ differs from $\phi_g$ by less than 0.1\%. Therefore, the geometric fill factor provides a sufficient approximation for estimating the fraction of photons contributing to drift current.

For the two detector geometries used in this work, the geometric fill factors were computed from direct optical measurements of the electrode structure. IDE-A had a geometric fill factor of $\phi_g^{\text{(A)}} = 0.513$ and IDE-B had $\phi_g^{\text{(B)}} = 0.586$.

At high bias voltages near or within the avalanche regime, the effective collection area may increase due to field fringing, enabling charge carriers generated just outside the nominal gap region to drift into the high-field zone. However, in this regime the detector response is dominated by carrier multiplication and any modest bias-dependent increase in $\phi_{\text{eff}}(V)$ is expected to be subdominant to the exponential gain behavior. Accordingly, the analysis in this work treats the fill factor as a geometric quantity, independent of voltage, which represents a conservative assumption since fringe effects can only increase the effective collection area.

\subsection*{Efficiency Metrics and Charge Extraction}
\label{sec:Metrics}

The number of charge carriers (holes) collected per pulse, \(n_{\text{h}}\), was determined from the peak amplitude of the fitted waveform. Given the nominal gain of the CR-110 charge-sensitive preamplifier, \(A = 1.4~\mathrm{V/pC}\), the measured voltage can be converted to charge and then to carrier number using the elementary charge \(e = 1.602 \times 10^{-19}~\mathrm{C}\):
\begin{equation}
n_{\text{h}} = \frac{V_{\text{peak}}}{A \cdot e}.
\end{equation}
This method assumes that charge is collected on a timescale much shorter than the preamplifier decay constant \(\tau = 140~\upmu\mathrm{s}\), so that peak amplitude accurately reflects the total collected charge. Charge transit time \(t_{\text{tr}}\) was estimated from
\begin{equation}
t_{\text{tr}} = \frac{L}{\mu_{h} E},
\end{equation}
where \(L\) is the transport distance, \(\upmu_{h}\) is the effective hole mobility, and \(E\) is the applied electric field. Although \(\upmu_{h}\) decreases at lower temperatures, it increases strongly with field, reaching values above \(0.01~\mathrm{cm^2/(V\cdot s)}\) at fields exceeding \(70~\mathrm{V/\upmu m}\), yielding sub-microsecond transit times~\cite{Hijazi2014}. At lower fields, where transit times increase, fitted rise times remained below \(20~\upmu\mathrm{s}\), well within the linear regime of the preamplifier response. This confirms that peak amplitude is a valid proxy for collected charge across all datasets.

To quantify detector response, external quantum efficiency (EQE) was defined as the ratio of collected charge carriers to the total number of incident photons $N_{\text{ph}}$:
\begin{equation}
\eta = \frac{n_{\text{h}}}{N_{\text{ph}}}.
\label{eq:eta_definition}
\end{equation}
This quantity includes all optical losses, geometric effects and reflects the overall performance of the device under test.

Only a subset of the incident photons are absorbed in field-active regions of the a-Se layer. To account for this, an effective photon count is defined:
\begin{equation}
N_{\text{eff}} = \phi_{g} \cdot N_{\text{ph}},
\end{equation}
where \(\phi_{g}\) is the geometric fill factor. This represents the number of photons incident on regions that contribute to charge collection.

From this, an intrinsic efficiency is defined:
\begin{equation}
\eta_{\text{int}} = \frac{n_{\text{h}}}{N_{\text{eff}}} = \frac{\eta}{\phi_{g}},
\end{equation}
which isolates the material and device response from purely geometric effects. Intrinsic efficiency reflects the probability of charge collection per photon in the active region, independent of detector layout. 

\subsection{Low-Photon Threshold Response at 401 nm}
\label{sec:LowP}

The sensitivity of a-Se detectors at cryogenic temperatures was evaluated using a matched filter algorithm designed to isolate low-amplitude signals in high-noise environments~\cite{Turin1960}. The matched filter output was used to quantify detection performance in terms of the probability of correctly identifying true signals and the trade-off with false positives, as summarized by the Receiver Operating Characteristic (ROC) curve. Detection efficiency is defined as the ratio of correctly identified signal events to the total number of signal events present in the data.

The output of the matched filter was obtained by convolving each noise-filtered waveform with a normalized, time-reversed template representing the detector response, as prescribed by matched filter theory. Mathematically, the matched filter output \( M(t) \) is expressed as:
\[
M(t) = \int_{t_0}^{t_1} V_{\text{filtered}}(\tau)\, T(t_1 - \tau)\, d\tau
\]
where \( V_{\text{filtered}} \) is the low-pass-filtered input waveform, \( T \) is the normalized template and \( \tau \) is the integration variable. The integration limits \( t_0 \) and \( t_1 \) were chosen to encompass the modeled signal onset and decay region. The template was generated by averaging a subset of filtered waveforms and normalizing the result over the fitted signal window. A 150\,kHz low-pass frequency-domain filter was applied to suppress high-frequency noise.

The integration window was defined using the waveform model introduced in Section~\ref{sec:WaveformModel}. The optimal interval was determined from the fitted parameters of the average waveform, specifically from the region extending from the fitted peak time \( x_0 \) to \( x_0 + n_{\tau}\, t_2 \), where \( t_2 \) is the decay constant. A systematic study varying \( n_{\tau} \), the scaling factor applied to \( t_2 \), established that \( n_{\tau} = 2.5 \) yielded the highest detection efficiency, corresponding to a window that captures approximately 68\% of the pulse area. A secondary optimization determined that a detection threshold of \( \upmu_n + 4\,\sigma_n \) provided an effective balance between rejecting noise and preserving true signals. Here, \( \upmu_n \) and \( \sigma_n \) are the mean and standard deviation of matched filter scores computed from the pre-pulse noise baseline. To confirm proper alignment, the matching window was shifted from $-200~\upmu\text{s}$ to $+200~\upmu\text{s}$ and detection efficiency was found to peak at zero offset, consistent with the signal onset.

ROC curve analysis was used alongside detection efficiency to evaluate classifier performance across a range of thresholds. The ROC curve is a parametric plot of true positive rate (TPR) versus false positive rate (FPR), and the area under the ROC curve (AUC) provides a scalar summary of performance, with \( \text{AUC} = 1 \) indicating perfect discrimination and \( \text{AUC} = 0.5 \) corresponding to random guessing.

\begin{figure}[H]
    \centering
    \includegraphics[width=0.5\textwidth]{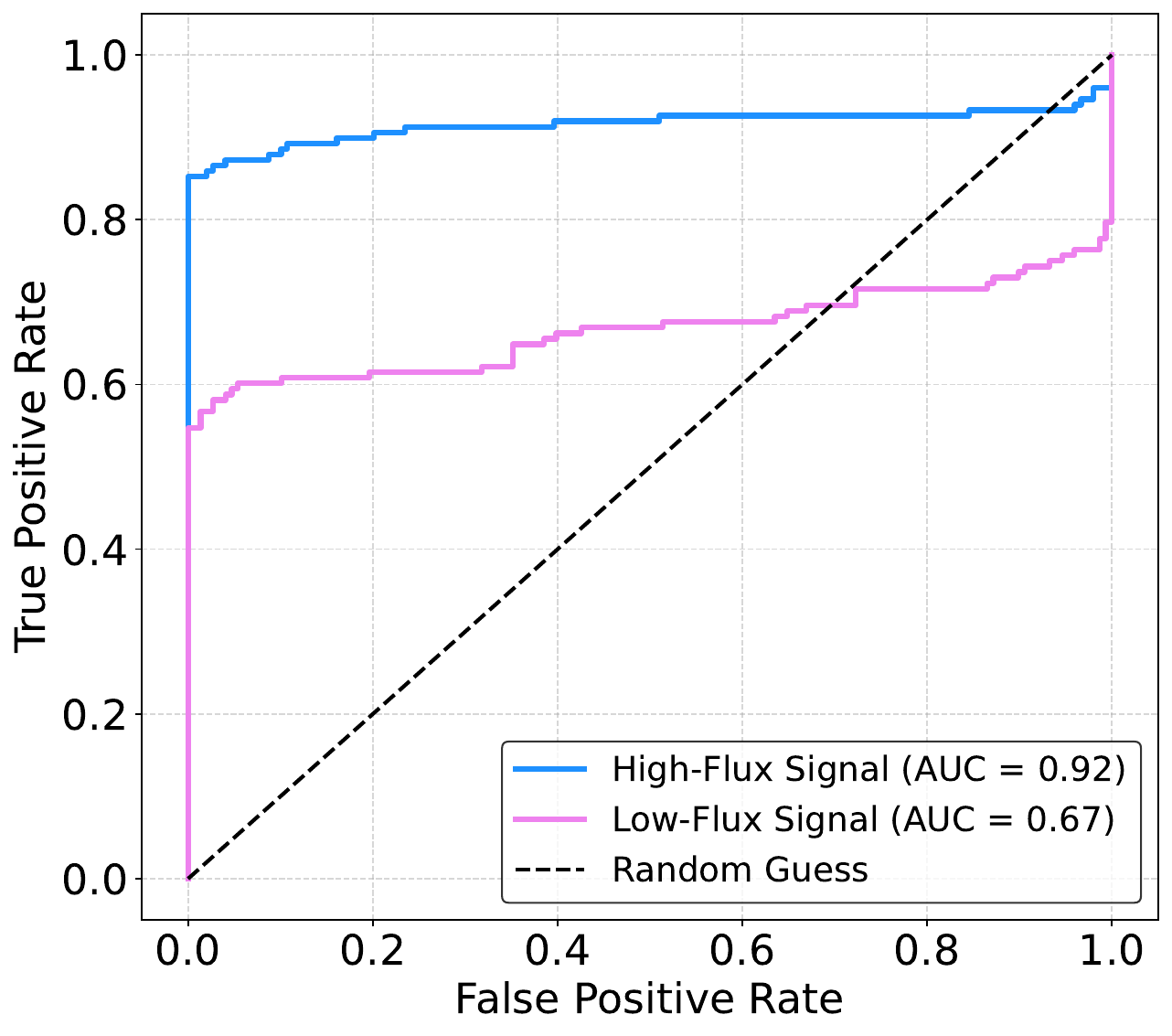}
    \caption{Matched filter ROC curves for two representative flux conditions. A high-flux signal (6816 photons at 87\,K) yields an AUC of 0.92. A low-flux signal (196 photons at 165\,K) results in an AUC of 0.67.}
    \label{fig:roc}
\end{figure}

Figure~\ref{fig:roc} compares ROC curves for two data points. The higher-flux condition, corresponding to 6816 incident photons at 87\,K, results in an AUC of 0.92 and a false positive rate of 0.1 at a true positive rate of 0.85. The corresponding detection efficiency was 80\%. In contrast, the lower-flux case at 196 photons and 165\,K yielded an AUC of 0.67 and a high false positive rate of 0.905 at TPR = 0.75, with efficiency reduced to 43.9\%. These two cases were selected as reference benchmarks to define a practical boundary for single-shot detection. Signals with AUC \( \geq 0.85 \) and efficiency \( \geq 0.8 \) were considered reliably detectable without averaging.

\begin{figure}[htbp]
    \centering
    \includegraphics[width=0.9\textwidth]{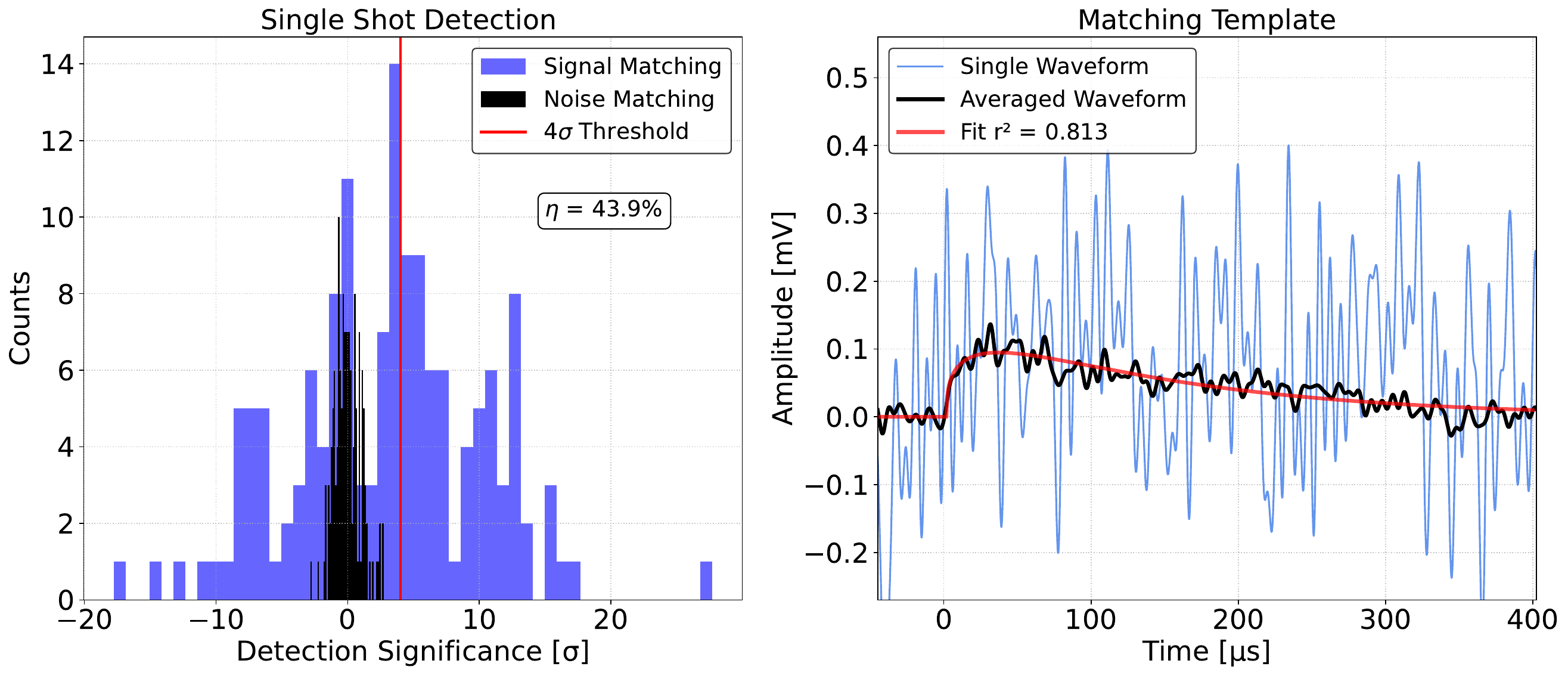}
    \caption{Matched filter performance for a low-flux event (196 photons at 165 K). Left: Histogram of matched filter scores for signal (violet) and noise (gray). The vertical line marks the 4$\sigma$ threshold. Only the violet bins to the right contribute to the measured detection efficiency of 43.9\%, indicating limited separability at low photon flux. Right: Template construction from averaged waveform and model fit.}
    \label{fig:matched1}
\end{figure}

\begin{figure}[htbp]
    \centering
    \includegraphics[width=0.9\textwidth]{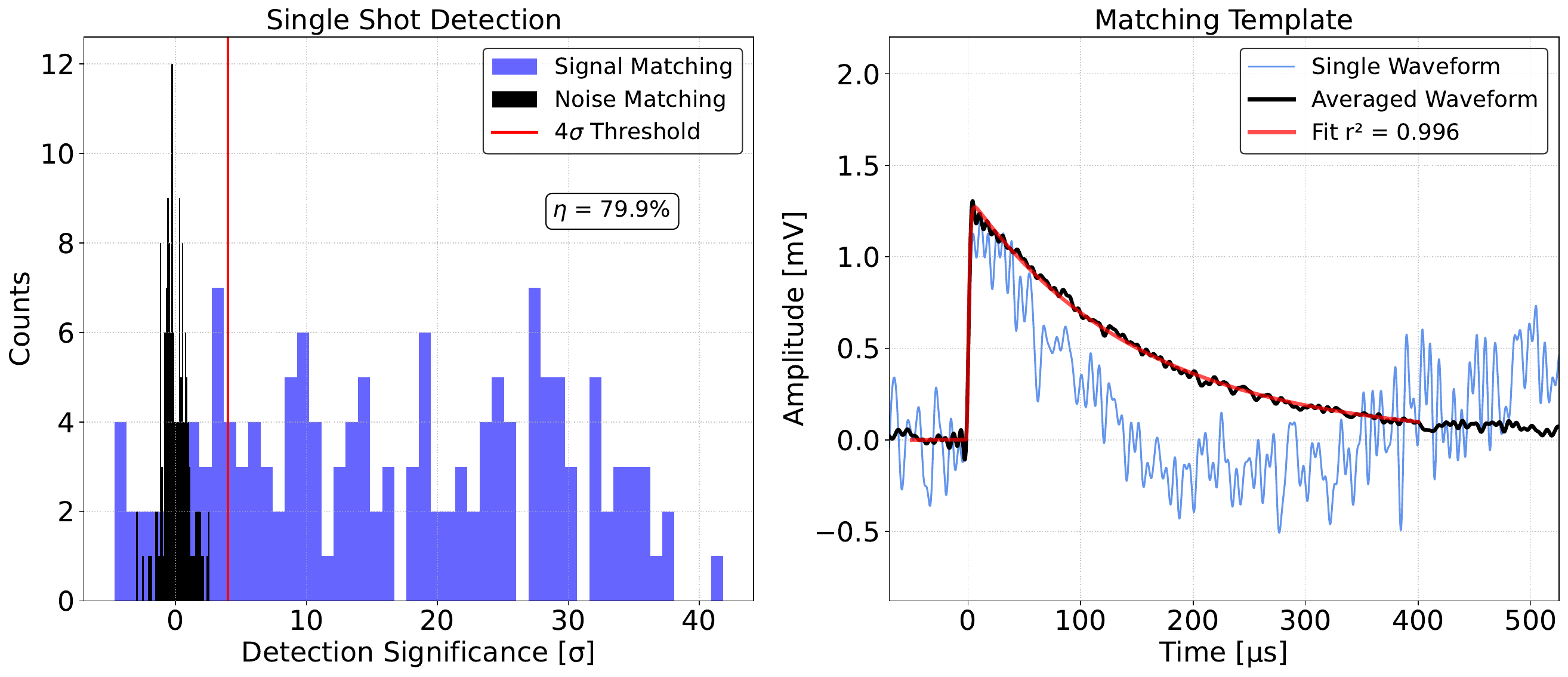}
    \caption{Matched filter performance for a higher-flux event (6816 photons at 87 K). Left: Matched filter score histogram shows strong separation between signal and noise, with most signal events exceeding the 4$\sigma$ threshold leading to a measured detection efficiency of 79.9\%. Right: Template construction from averaged waveform and model fit.}
    \label{fig:matched2}
\end{figure}

Figures~\ref{fig:matched1} and \ref{fig:matched2} show the matched filter output distributions and template fits for the low flux and high flux cases. At high flux, signal and noise responses are well separated, while at low flux the overlap is significant. The matching template was derived from FFT-filtered, averaged waveforms using the waveform modeling and fit procedure described in Section~\ref{sec:WaveformModel}. In the low flux case, Figure~\ref{fig:matched1} left, signal and noise distributions overlap significantly. Only a small fraction of signal events exceed the 4$\sigma$ threshold, and these are counted as successful detections. In contrast, the high flux case, Figure~\ref{fig:matched2} left, shows strong separation, with most signal scores clearly above the threshold and noise scores remaining below it.

To validate the match interval, the matching window was shifted from $-200~\upmu\text{s}$ to $+200~\upmu\text{s}$. Detection efficiency peaked at zero offset, confirming the alignment procedure and selected window were appropriate.

\begin{figure}[htbp]
    \centering
    \includegraphics[width=\textwidth]{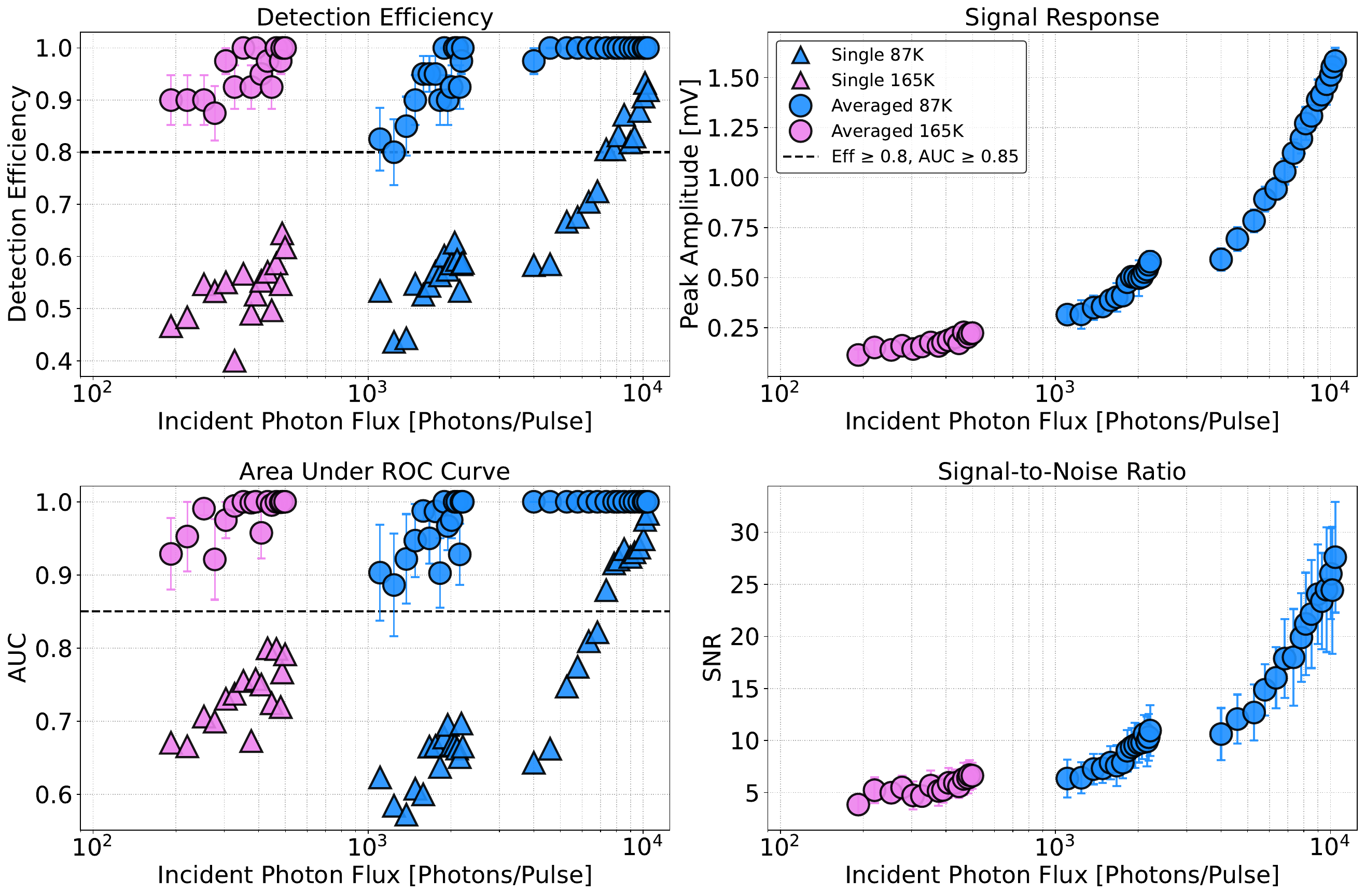}
    \caption{Matched filter performance versus incident photon flux. Top left: Detection efficiency for single waveforms (triangles) and 12-waveform averaging (circles). Top right: Average response amplitude from fit. Bottom left: AUC for single and averaged detection. Bottom right: SNR\textsubscript{M} computed from the matched region. Benchmark thresholds at \(\eta = 0.8\) and AUC = 0.85 are shown with dashed lines.}
    \label{fig:lowquad}
\end{figure}

Figure~\ref{fig:lowquad} summarizes the matched filter performance across three datasets acquired at fields corresponding to avalanche operation: a low-flux set (192--499 photons) at 165\,K and 110.7\,V/\textmu m, a middle-flux set (1105--2209 photons) at 87\,K and 119.2\,V/\textmu m and a high-flux set (4003--10405 photons) at 87\,K and 119.2\,V/\textmu m. The matched filter detection efficiency increases with photon flux for both single-shot and averaged signals. When twelve-waveform averaging was applied to all datasets, performance exceeded the defined AUC and efficiency benchmark thresholds across all flux levels, including the lowest-flux case. The matched filter signal-to-noise ratio, \( \text{SNR}_{\text{M}} \), used here to characterize detection performance, was computed as
\[
\text{SNR}_{\text{M}}  = \frac{A_{\text{peak}}}{\sigma_{\text{pre}}}
\]
where \( A_{\text{peak}} \) is the peak matched filter output and \( \sigma_{\text{pre}} \) is the standard deviation of the pre-pulse baseline region from \(-350\) to \(-50\,\upmu\text{s} \). This definition reflects a peak-over-noise ratio specific to single-shot matched filter outputs and differs from power-based SNR metrics used in the waveform fitting method described in Section~\ref{sec:WaveformModel}. Both the average waveform response (top right) and \( \text{SNR}_{\text{M}} \) (bottom right) scale with photon flux.

At 87 K, the single-shot performance is reduced relative to 165 K. At fields below avalanche, a negative temperature dependence in photoresponse in a-Se is attributed to decreasing mobility–lifetime product \cite{Tsuji1994,Kasap2015}. At higher fields, like those used in the present data where avalanche multiplication occurs, the signal amplitude also decreases with decreasing temperature due to reduced impact ionization rates \cite{Tsuji1994}.

The minimum SNR\textsubscript{M} observed was 3.8 in the lowest flux case, increasing to 27.6 at the highest flux. Averaging improved both SNR\textsubscript{M} and AUC, particularly in the low-flux regime below 1000 photons, where single-shot detection performance is limited by noise. Single-shot performance began to plateau near 6800 photons, where the AUC and efficiency crossed their respective thresholds. These results define the practical sensitivity limits of the detector at 87\,K and 165\,K 
with the current design of detector and the readout electronics. A lower single-shot performance seems achievable with better SNR\textsubscript{M}, specifically in the readout electronics. 

While performance is reported in terms of the total number of photons incident on the a-Se dot area of IDE-A, as shown in Figure~\ref{fig:dots}, only a subset are absorbed within field-active regions of the device. As discussed in Section~\ref{sec:Metrics}, using the geometric fill factor $\phi_g^{\text{(A)}} = 0.513$, the corresponding field-effective photon counts for the benchmark cases are approximately 100 and 3400 photons, respectively. These values reflect the number of photons incident on the lateral gap region where drift and collection are possible and more directly represent the true excitation load seen by the detector. When expressed in these terms, the low-flux benchmark corresponds to just 100 effective photons, further underscoring the sensitivity of the a-Se photoconductor to weak, localized excitation. As future designs seek to improve active area coverage with blocking layer optimization, the field-effective photon count provides a meaningful basis for quantifying low-photon detection thresholds demonstrated here in lateral a-Se devices operating in avalanche.

\subsection{Linearity and High-Field Response at 93 K}
\label{sec:linearity_response}

\begin{figure}[htbp]
    \centering
    \includegraphics[width=1\textwidth]{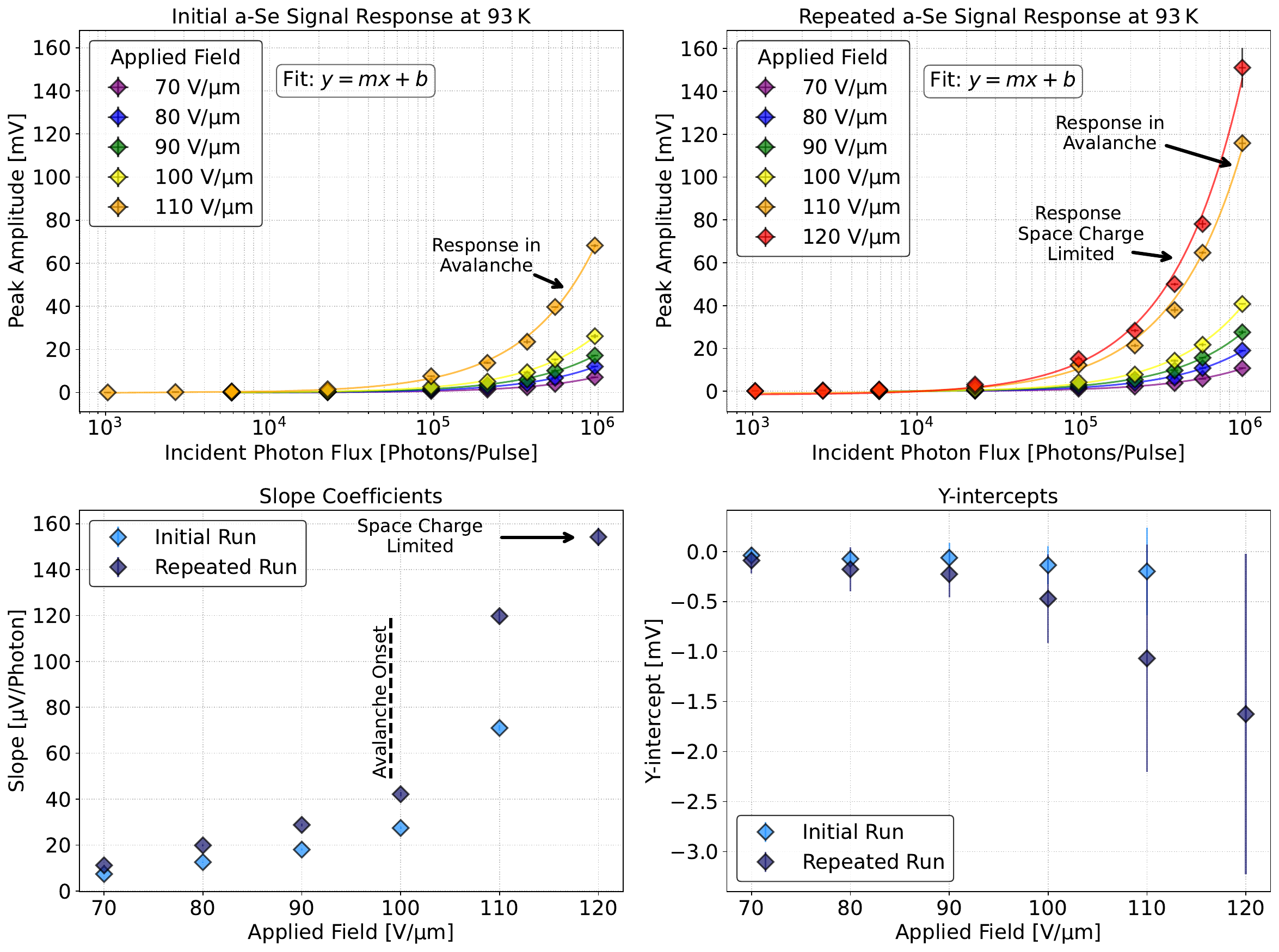}%
    \caption{Response of the IDE-A a-Se detector at 93\,K. Top: Peak amplitude versus incident photon flux for the initial run (left) and repeated run (right) over a range of applied fields. Statistical errors are included but too small to be visible. The $x$-axis is logarithmic to accommodate the wide dynamic range in photon flux. Linear fits \( y = mx + b \) are overlaid to quantify the response at each field; avalanche and space-charge-limited responses are indicated. Bottom: Extracted slope coefficients (left) and $y$-intercepts (right) as a function of field for both runs.}
    \label{fig:response_plots}
    
\end{figure}

The IDE-A sample was tested at 93\,K across a wide range of incident photon fluxes and applied electric fields to evaluate linearity and high-field behavior. Two datasets were acquired: an initial run and a repeated run performed two days later under identical conditions. The results are shown in Figure~\ref{fig:response_plots}, which plots peak amplitude as a function of incident photon flux for each field and filter stack configuration A-I in Table~\ref{tab:filter_stacks}. All measurements were performed with the detector biased between 70 and 120\,V/\textmu m. The horizontal axis is presented on a logarithmic scale to accommodate the large dynamic range in photon input. For each measurement point, the peak amplitude was determined by averaging nine waveforms, yielding an SNR as defined in Section~\ref{sec:WaveformModel} above 8 dB at the lowest photon flux.

In both datasets, the detector response was approximately linear at moderate fields, with a clear change in slope emerging as the applied field increased beyond 100 V/$\upmu$m, marking the onset of avalanche multiplication. By 110 V/$\upmu$m, the avalanche behavior is visibly pronounced in both the response curves and the slope coefficient plot. This onset field is consistent with the avalanche gain observed near 101 V/$\upmu$m in the comparative measurements presented in Section~\ref{sec:Te_aSe_comparison}, as shown in the left panel of Figure~\ref{fig:teaSe} and summarized in Table~\ref{tab:model_parameters}. The repeated run includes an additional data set at 120\,V/\textmu m, which shows a further increase in amplitude, though less than proportional to the field increase. This sublinear behavior is attributed to space-charge-limited (SCL) gain, where internal field screening, potentially due to accumulated avalanche carriers, suppresses further charge amplification. This regime, labeled in the plot, is also evident in the VUV response at 87\,K in the right panel of Figure~\ref{fig:vuvpllot} in Section~\ref{sec:direct_vuv} and in the response curves at 93\,K and 297\,K in the left panel of Figure~\ref{fig:teaSe} in Section~\ref{sec:Te_aSe_comparison}.

The extracted slope coefficients for both runs are shown in the lower left panel of Figure~\ref{fig:response_plots}. At each overlapping field, the repeated run exhibits enhanced response relative to the initial run. Using a symmetrized percent difference, the repeated run slopes were higher by 41--51\%, with the largest deviation occurring at 110\,V/\textmu m. This enhancement may reflect changes in trap occupancy, relaxation of contact injection conditions, or minor thermal variations. A detailed transport-based comparison is deferred to Section~\ref{sec:Te_aSe_comparison}, where these results are discussed alongside a-SeTe response and model trends.

The lower right panel shows the corresponding y-intercepts from the linear fits. These values are negative in both datasets and trend further from zero with increasing field. The repeated run shows more negative intercepts than the initial run, especially above 100\,V/\textmu m. The y-intercept reflects the extrapolated detector response at zero photon flux, which becomes increasingly negative at high fields potentially due to field-induced baseline shifts and space-charge effects discussed earlier. 

These measurements demonstrate that lateral a-Se devices maintain a stable and approximately linear response over a wide dynamic range of photon flux at 93 K. The avalanche behavior in this configuration further underscores the capability of lateral devices to operate in high-field regimes at cryogenic temperature.

\subsection{Direct Detection of Narrowband 130 nm VUV Light}
\label{sec:direct_vuv}

To evaluate direct detection of liquid noble gas scintillation light, an IDE-B detector was exposed to pulsed VUV illumination from an argon flash lamp with an emission peak at 130 $\pm$ 2.5 nm, a full-width at half-maximum of 20\,nm ± 5\,nm and a peak transmission of 15\%. The entire interdigitated region was coated with a-Se to maximize the photosensitive area and reduce possible photoelectric effects from uncoated gold electrodes.

\begin{figure}[htb]
    \centering
    \includegraphics[width=0.93\textwidth]{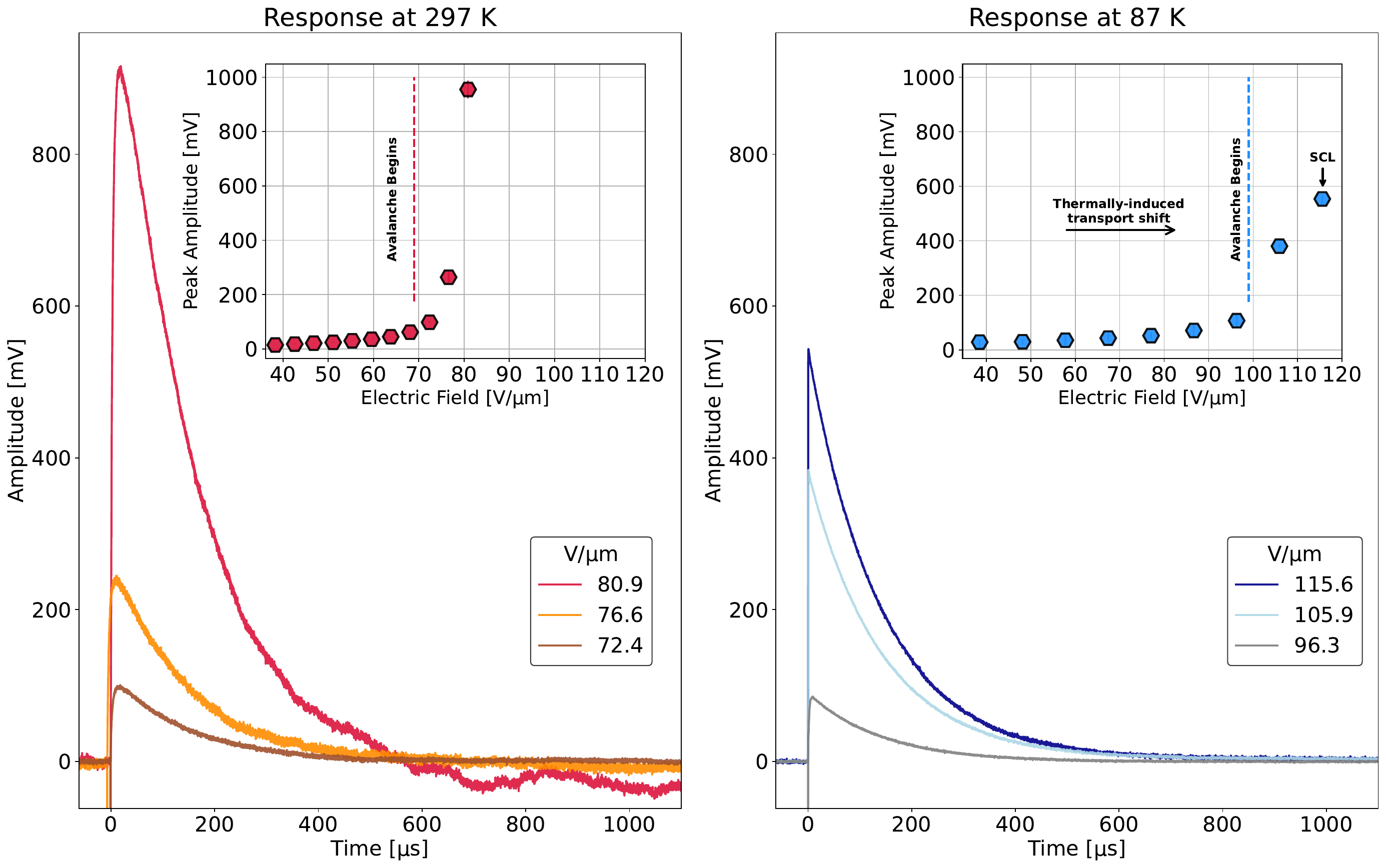}%
    \caption{Response of an IDE-B detector with full-area a-Se coating to pulsed 130\,$\pm$\,2\,nm VUV light from an argon flash lamp. Left: Single waveforms at 297\,K in the avalanche regime. Right: Corresponding single waveforms at 87\,K. Insets show mean peak amplitude versus applied field, with avalanche onset marked and SCL indicated in the 87 K response. Statistical errors in the insets are included but too small to be visible.}
 
\label{fig:vuv_response}
    \label{fig:vuvpllot}
\end{figure}

Figure~\ref{fig:vuv_response} shows representative waveforms from both room temperature (297\,K, left) and cryogenic (87\,K, right) operation. In each case, three signals in the avalanche regime are shown. The onset of avalanche differs between the two datasets: near 68\,V/\textmu m at 297\,K and near 99\,V/\textmu m at 87\,K, as indicated in the inset plots. This temperature-dependent shift in avalanche onset has been consistently observed across lateral a-Se devices in this work, as discussed in Section~\ref{sec:Te_aSe_comparison}. The present data demonstrate that this trend persists for deep ultraviolet photoexcitation.

The mean peak amplitudes shown in the insets were determined by approximately 20 waveforms at each field using the waveform fitting method described in Section~\ref{sec:WaveformModel}. Peak amplitudes were extracted from the fits and averaged, with statistical errors included in the plot but too small to be visible.

While a full transport analysis is beyond the scope of this work, the observed shift in avalanche onset field is consistent with thermally-activated hole mobility and field-assisted gain. Similar behavior is discussed in Section~\ref{sec:Te_aSe_comparison} for a-SeTe and in Section~\ref{sec:linearity_response} for cryogenic high-field response.

As noted in Section~\ref{sec:ArLamp}, the photon flux of the argon flash lamp was not calibrated, preventing determination of an absolute detection efficiency. The optical absorption of a-Se in the VUV is substantially higher than in the visible, leading to an expected detector response at 130 nm that exceeds that at 401 nm \cite{Leiga1968}. Additionally, Onsager dissociation probability also increases with photon energy, resulting in higher photogeneration efficiency at shorter wavelengths \cite{Pai1975}.

These results confirm that a-Se remains photosensitive to VUV light at cryogenic temperature and can operate in the avalanche regime at field strengths compatible with low-temperature environments, providing evidence of direct VUV-driven avalanche response in a lateral a-Se device.

\subsection{Tellurium-Doped a-Se Response Comparison}
\label{sec:Te_aSe_comparison}

Tellurium-doped a-Se detectors were tested under the same temperatures and applied electric fields as the undoped devices to assess the impact of material variation on detector performance and to establish that avalanche gain is achievable at cryogenic temperatures. The field-dependent quantity \(\eta(E)\), defined in Eq.~\ref{eq:eta_definition}, represents the external quantum efficiency. The measured response curves exhibit four distinct regions: trap-limited (TL) where carrier transport is dominated by trapping and detrapping, saturation-limited (SL) where traps fill and mobility saturates, avalanche (AV) where impact ionization produces gain and space-charge-limited (SCL) where carrier accumulation limits further current increase.  

Each curve is analyzed using a composite model function \(\eta(E; k_1, n_1, k_2, n_2, A, b, x_0)\) that captures the first three transport regions observed in the data:

\begin{equation}
\eta(E) =
\begin{cases}
k_1 E^{n_1} & \text{for } E < E_{\text{TL}\to\text{SL}} \quad \text{(trap-limited)} \\[10pt]
k_2 E^{n_2} + \Delta_1 & \text{for } E_{\text{TL}\to\text{SL}} \leq E < E_{\text{SL}\to\text{AV}} \quad \text{(saturation-limited)} \\[10pt]
A \exp\left[\left(\dfrac{E}{x_0}\right)^b\right] + \Delta_2 & \text{for } E \geq E_{\text{SL}\to\text{AV}} \quad \text{(avalanche)}
\end{cases}
\label{eq:composite_model}
\end{equation}

Continuity is enforced using the additive terms
\begin{align}
\Delta_1 &= k_1 E_{\text{TL}\to\text{SL}}^{n_1} - k_2 E_{\text{TL}\to\text{SL}}^{n_2} \\[6pt]
\Delta_2 &=
\begin{cases}
\eta(E_{\text{SL}\to\text{AV}}^-) - A \exp\left[\left(\dfrac{E_{\text{SL}\to\text{AV}}}{x_0}\right)^b\right] & \text{if SL region is present} \\[6pt]
k_1 E_{\text{SL}\to\text{AV}}^{n_1} - A \exp\left[\left(\dfrac{E_{\text{SL}\to\text{AV}}}{x_0}\right)^b\right] & \text{otherwise}
\end{cases}
\end{align}

The choice of this empirical form is based on the observation that all datasets display extended linear regions in log-log space. While physical transport models such as Poole–Frenkel can describe low-field behavior, they are limited in field range and did not provide satisfactory fits. In contrast, this model captures the broad transition from trap-limited to avalanche behavior and reflects the observed field response without assuming a particular physical transport mechanism that cannot account for the full field range.

\begin{figure}[H]
    \centering
    \includegraphics[width=\columnwidth]{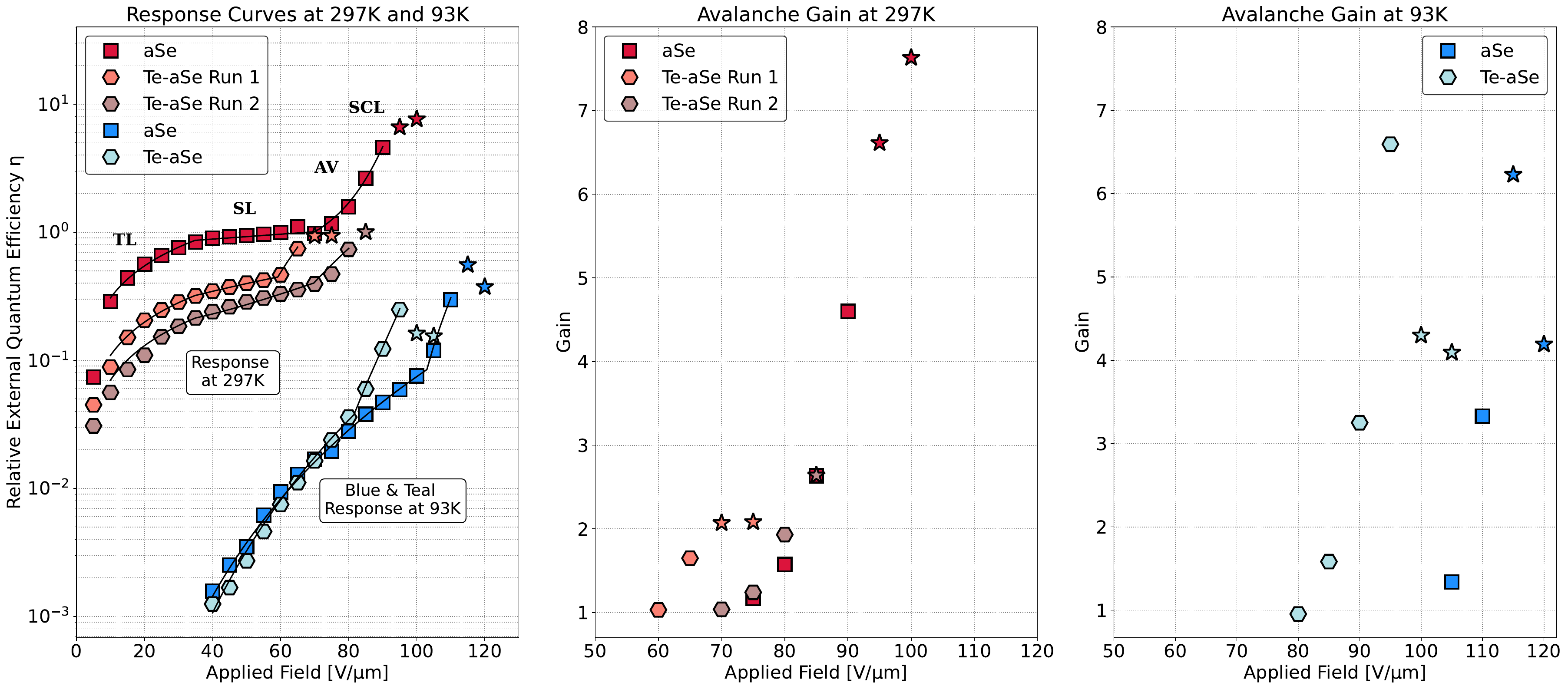}
    \caption{Field-dependent efficiency curves for a-Se and a-SeTe detectors at 297\,K and 93\,K. The left panel shows the a-Se response with two a-SeTe runs at 297\,K and a single response for each material at 93\,K. Labeled regions above the 297\,K curve indicate trap-limited  (TL), saturation-limited (SL), avalanche (AV) and space-charge-limited (SCL) behavior. SCL data are marked with stars. All efficiency data are scaled to the avalanche onset of the 297\,K a-Se response to illustrate relative trends across devices and temperatures. Each dataset is fit (solid black lines) using the composite model of Eq.~\ref{eq:composite_model}, with the 93\,K fits omitting the SL region. The middle and right panels show avalanche gain plotted against applied field, scaled to their respective avalanche onsets. The 297\,K a-Se device reaches a maximum gain of 4.6 (7.6 including SCL), while at 93\,K the a-SeTe detector exceeds the a-Se gain, reaching 6.6 compared to 6.2. Compared to undoped a-Se, the a-SeTe detectors exhibit earlier avalanche onset and higher gain at cryogenic temperature but lower quantum efficiencies at all temperatures tested.
}
    \label{fig:teaSe}
\end{figure}

Figure~\ref{fig:teaSe} shows the composite fits to the EQE data at 297\,K and 93\,K. All datasets are scaled to the avalanche onset of the 297\,K a-Se response, which exhibited the largest signal among temperatures studied. The 297\,K a-Se response, shown in the left panel, contains the full sequence of transport behaviors labeled as TL, SL, AV and SCL. These regions are labeled above the 297\,K curve to serve as a visual reference for interpreting the other datasets. Two a-SeTe detectors were tested at this temperature. One shows an earlier onset of avalanche at 59.2\,V/\textmu m compared to the 70\,V/\textmu m observed in both the a-Se device and the second a-SeTe run. The reduced response amplitude and variation between nominally identical detectors suggest possible differences in composition or morphology.

At 93\,K both materials exhibit high-field avalanche behavior but do not display a well-defined saturation-limited region. Fits to these curves use only a single power law followed by the exponential avalanche term. The a-SeTe detector shows earlier avalanche onset at 81.5\,V/\textmu m compared to 101\,V/\textmu m for a-Se. This measurement confirms that avalanche occurs in a-SeTe at cryogenic temperatures and at lower electric fields.

To isolate high-field behavior, the gain is plotted relative to its value at the onset of avalanche for 297\,K in the center panel and 93\,K in the right panel. The 297\,K a-Se device reaches a maximum gain of 4.8, increasing to 7.8 when including points in the SCL regime. At 93\,K the a-SeTe sample achieves a gain of 6.8 while the a-Se device reaches 6.2. This gain enhancement despite the lower onset field suggests that the doped material remains active under cryogenic operation.

Table~\ref{tab:model_parameters} reports the fitted composite model parameters for each dataset at 297\,K and 93\,K. These parameters reflect the overall scaling of the efficiency response and the underlying transport behavior in each material and temperature. Fits were constrained to remain continuous at the transition fields. The observed trends in the power-law exponents and avalanche fields demonstrate the evolution of carrier collection with temperature and confirm that a-SeTe exhibits avalanche behavior across the full temperature range tested, showing that both doped and undoped lateral a-Se devices sustain avalanche operation at cryogenic temperature. Compared to undoped a-Se, tellurium doping lowers the avalanche onset field by approximately 15–20\,V/$\upmu$m, achieves comparable or slightly higher gain at cryogenic temperature, but reduces both external and intrinsic quantum efficiencies, suggesting increased trap-assisted losses in the doped material \cite{Hellier2023}.

\begin{table}[ht]
\centering

\begin{tabular}{lccc}
\toprule
Temperature & Device    & $\eta$ at Onset & $\eta_{\text{int}}$ at Onset \\
\midrule
\phantom{0}297\,K     & a-Se       & 0.513   & 0.999 \\
                     & a-SeTe Run 1   & 0.209   & 0.408 \\
                     & a-SeTe Run 2   & 0.176   & 0.343 \\
\addlinespace
\phantom{0}93\,K      & a-Se       & 0.041   & 0.081 \\
                     & a-SeTe   & 0.017   & 0.034 \\
\bottomrule
\end{tabular}
\caption{External and intrinsic quantum efficiency at the onset of avalanche. Both materials exhibit a similar proportional decrease in $\eta_{\text{int}}$ when cooled, suggesting comparable temperature dependence of intrinsic efficiency.}

\label{tab:eff}
\end{table}

Table~\ref{tab:eff} summarizes the external and intrinsic quantum efficiencies discussed in Section~\ref{sec:Metrics} at the onset of avalanche for all devices. At 297\,K the intrinsic efficiency of undoped a-Se reaches 99.9\%, indicating that nearly all photons incident on field-accessible regions result in collected holes. The external efficiency \(\eta = 0.513\) matches the geometric fill factor \(\phi_{g} = 0.513\), implying that charge collection within the active region is nearly complete. This close agreement indicates that remaining losses arise primarily from geometric exclusion rather than material or interfacial effects. Tellurium-doped variants exhibit reduced intrinsic efficiency, suggesting increased trap-assisted losses, though avalanche multiplication remains observable. At 93\,K the intrinsic efficiency \(\eta_{\text{int}}\) drops significantly, reflecting suppressed carrier transport under cryogenic conditions. Since \(\eta_{\text{int}} = \eta / \phi_{g}\), this decrease directly follows from the reduced external quantum efficiency \(\eta\), which reflects the fraction of carriers collected in the field-active region. These results emphasize the critical role of field strength and material composition in maintaining charge collection efficiency at low temperature. Compared to undoped a-Se, tellurium doping lowers the avalanche onset field by approximately 15–20\,V/$\upmu$m, achieves comparable or slightly higher gain at cryogenic temperature, but reduces both external and intrinsic quantum efficiencies, consistent with increased trap-assisted losses in the doped material.

\begin{table}[ht]
\centering
\footnotesize
\begin{tabular}{|c|c|c|c|c|c|c|c|c|c|c|}
\hline
Temp & Device 
& $k_1$ & $n_1$ & $k_2$ & $n_2$ & $x_0$ & $b$ & $A$ 
& \multicolumn{1}{c|}{TL$\to$SL} & \multicolumn{1}{c|}{SL$\to$AV} \\
& & \scriptsize{$(V/\upmu\mathrm{m})^{-n_1}$} & & \scriptsize{$(V/\upmu\mathrm{m})^{-n_2}$} & & \scriptsize{V/$\upmu$m} & & & \scriptsize{V/$\upmu$m} & \scriptsize{V/$\upmu$m} \\
\hline
297 & a-Se            & 4.603e-2   & 0.827 & 1.951e-3  & 1.152 & 78.0  & 5.63  & 0.480 & 35.0  & 70.0  \\
297 & a-SeTe 1        & 1.251e-2   & 0.919 & 8.664e-1  & 0.153 & 64.1  & 2.46  & 0.589 & 36.7  & 59.1  \\
297 & a-SeTe 2        & 4.233e-3   & 1.107 & 2.623     & 0.070 & 74.1  & 7.98  & 0.078 & 36.8  & 65.0  \\
93  & a-Se            & 1.825e-10  & 4.306 & --       & --    & 106   & 7.24  & 0.153 & --    & 101.0 \\
93  & a-SeTe          & 1.058e-11  & 5.004 & --       & --    & 89.3  & 6.52  & 0.079 & --    & 81.6  \\
\hline
\end{tabular}
\caption{Fit parameters for the composite model curves shown in Figure~\ref{fig:teaSe} (left panel) for 297\,K and 93\,K response data. At 93\,K, the fits use a reduced model omitting the SL region due to the absence of a TL$\to$SL transition. The reduced $\chi^{2}$/dof values for each fit are: 0.96 for a-Se at 297\,K, 0.93 for a-SeTe Run~1 at 297\,K, 0.80 for a-SeTe Run~2 at 297\,K, 1.25 for a-Se at 93\,K and 1.20 for a-SeTe at 93\,K.}
\label{tab:model_parameters}
\end{table}

\section{Conclusion}\label{sec:conclusion}

The lateral a-Se photodetectors characterized in this study maintained stable operation from 297~K down to 87~K, with electronic noise quantified as a mean ENC of $2804 \pm 226~\mathrm{e^{-}}$ at 297~K and a gradual reduction observed at lower temperatures, as detailed in Section~\ref{sec:readout}. Matched-filter analysis shows that single-shot detection efficiencies $\geq 0.8$ with AUC $\geq 0.85$ are achieved with as few as $\sim\!6800$ incident 401\,nm photons at 87\,K, corresponding to approximately 3400 photons within field-active regions when accounting for the geometric fill factor. 

Across five decades of photon flux at 93\,K, the devices exhibit a linear response up to 100\,V/$\upmu$m, beyond which avalanche multiplication enhances the collected signal. Detectors also directly sensed narrow-band 130\,nm VUV light representative of liquid-argon scintillation at both 297\,K and 87\,K, confirming that a-Se remains VUV-responsive in the avalanche regime and that the onset field scales predictably with temperature.

Tellurium-doped a-Se devices reproduced the full transport sequence of undoped samples while lowering the avalanche threshold to 81\,V/$\upmu$m at 93\,K and attaining slightly higher gain, with a recorded peak of 6.6 compared to 6.2 for the undoped device at the same temperature. For the undoped device, the highest gain observed was 7.6 at 297\,K. In all cases, space-charge effects ultimately limited the maximum applicable field. Although the quantum efficiency at avalanche onset for the undoped device at 297\,K was $\eta = 0.513$, matching the geometric fill factor, this implies an intrinsic efficiency of $\eta_{\text{int}} = 0.999$, indicating nearly ideal charge collection within field-accessible regions. The a-SeTe devices showed reduced external efficiencies, with $\eta = 0.209$ and $0.176$, corresponding to $\eta_{\text{int}} = 0.408$ and $0.343$, respectively. Despite these losses, the earlier avalanche onset and enhanced gain at cryogenic temperature suggest that doping alters the carrier dynamics in a favorable manner, possibly reducing the required bias for high-gain operation. These results highlight the potential for compositional tuning to optimize performance in cryogenic photodetector applications.

Taken together, the results presented here demonstrate that lateral a-Se photodetectors combine low-photon sensitivity, wide dynamic range, cryogenic stability and direct VUV response in a lateral architecture. The detectors achieved single-shot detection efficiencies approaching 65\% with as few as 100 field-effective 401 nm photons at 165 K, illustrating their sensitivity to such low-level optical excitation. At 87\,K, the same architecture maintained comparable behavior under direct VUV illumination at 130\,nm, confirming that the avalanche response mechanism remains active at cryogenic temperature and across excitation wavelengths relevant to liquid-argon scintillation. The complementary measurements at 93\,K demonstrated linearity across five decades of photon flux before transitioning into a well-defined avalanche regime, further validating the stability of the transport processes. Finally, the comparison with tellurium-doped variants revealed that compositional tuning can reduce avalanche onset fields and enhance gain, suggesting pathways to improved performance through controlled doping.

Although the present devices do not achieve single-photon detection, the demonstrated performance reflects an unoptimized architecture rather than a fundamental material limitation. The detectors employed commercial IDE substrates with polyimide covering both electrodes, a configuration that suppresses hole injection but also reduces charge collection at the receiving electrode, limiting the observable avalanche gain. With optimized electrode designs that expose the collecting electrodes and incorporate selective blocking layers to prevent electron injection, significantly higher multiplication factors are expected. Prior studies using vertical a-Se structures have reported gains exceeding 100 at room temperature and over 1000 with compositional and contact engineering \cite{Reznik2007}. While impact ionization is reduced at cryogenic temperatures, this regime remains underexplored, and the present results show clear avalanche behavior persisting down to 87 K. Further reduction in electronic noise through optimized cryogenic front-end design could lower the detection threshold toward a few hundred photons per pulse. Based on these trends, the extrapolated path toward single-photon sensitivity appears physically attainable through concurrent improvements in device geometry, blocking layer design and readout electronics.

In light of these results, lateral a-Se photodetectors represent a compelling option for next-generation pixelated liquid-argon TPCs, offering scalable tiling, low radioactivity, high-field operation and VUV-sensitive photon detection. While a direct quantitative comparison of VUV and 401\,nm response on the same device was not performed, the observed temperature-dependent avalanche response of undoped a-Se remains consistent across devices and wavelengths. Under VUV excitation on IDE-B, the avalanche onset fields were 68\,V/$\upmu$m at 297\,K and 99\,V/$\upmu$m at 87\,K. Under 401\,nm excitation on IDE-A, the corresponding onset fields were 70\,V/$\upmu$m at 297\,K and 101\,V/$\upmu$m at 93\,K.

\section*{Acknowledgements}

This work was supported by the U.S. Department of Energy through the Office of Science Graduate Student Research (SCGSR) program under contract DE-SC0014664. The author gratefully acknowledges the support and resources provided by the Physics Division at Oak Ridge National Laboratory. Special thanks are extended to Dr. Mathieu Benoit for assistance with profilometer measurements and use of characterization equipment and to Toby King, Technical Professional in Experimental Physics, whose expertise and professionalism have been invaluable throughout this work. This material is also based upon work supported by the U.S. Department of Energy, Office of Science, Office of High Energy Physics under Award Number DE-SC0020065. Additional support was provided by the National Science Foundation under award number NSF-PHY 2110985 and U.S. Department of Energy, National Nuclear Security Administration under Award Number DE-NA0004066 (Rutgers) and from the Science and Technology Facilities Council (STFC) under grant number ST/W003945/1.

\appendix
\section{SiPM Calibration}
\label{sec:sipmcompare}
\begin{figure}[htb]
    \centering
    \includegraphics[width=\textwidth]{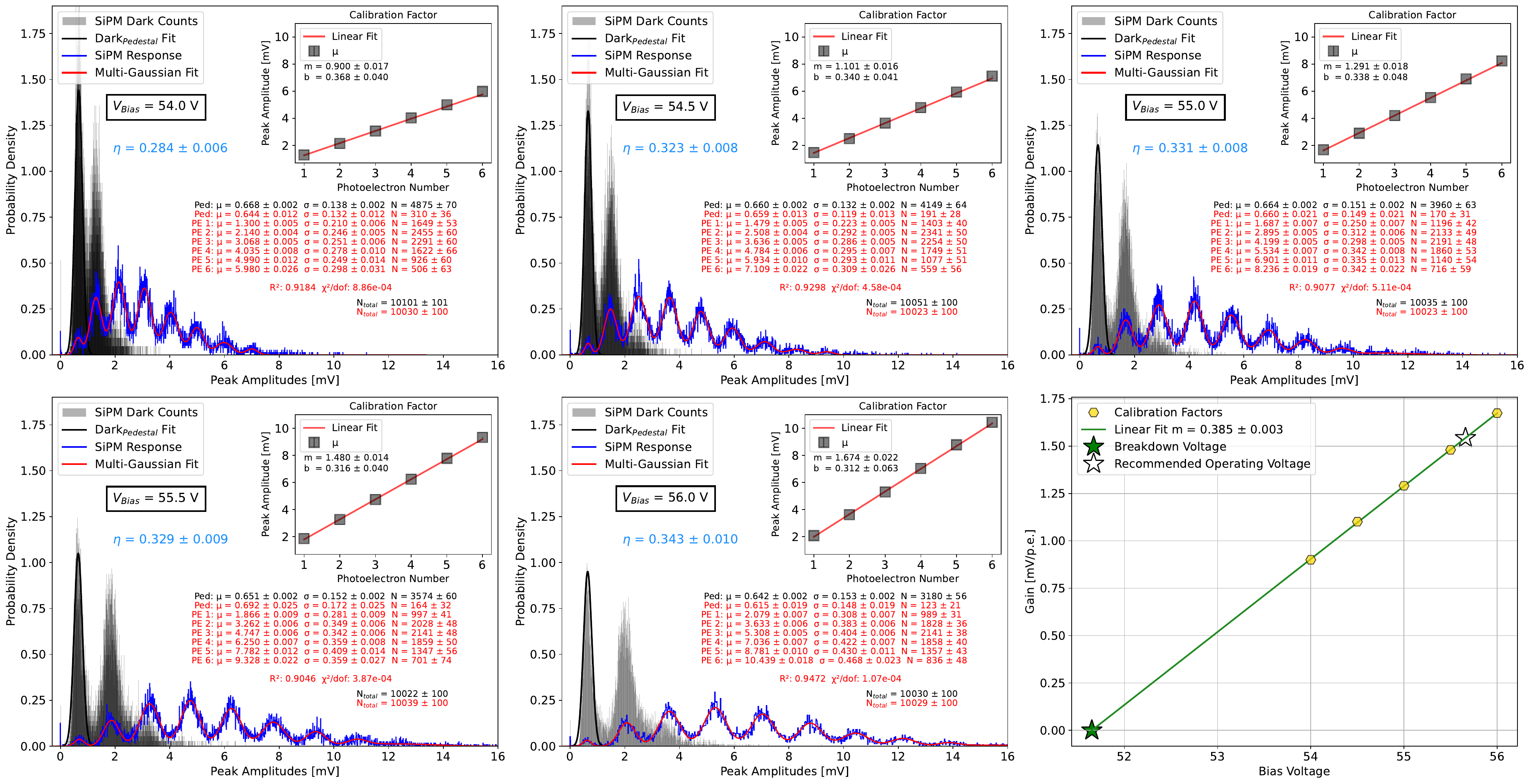}
    \caption{Summary of SiPM gain calibrations from 54\,V to 56\,V. Subplots show representative multi-Gaussian fits for each bias voltage with a final panel displaying the extracted gain versus bias. Breakdown and recommended operating voltages are indicated.}
    \label{fig:sipmcal}
\end{figure}

\begin{figure}[htbp]
    \centering
    \includegraphics[width=0.8\textwidth]{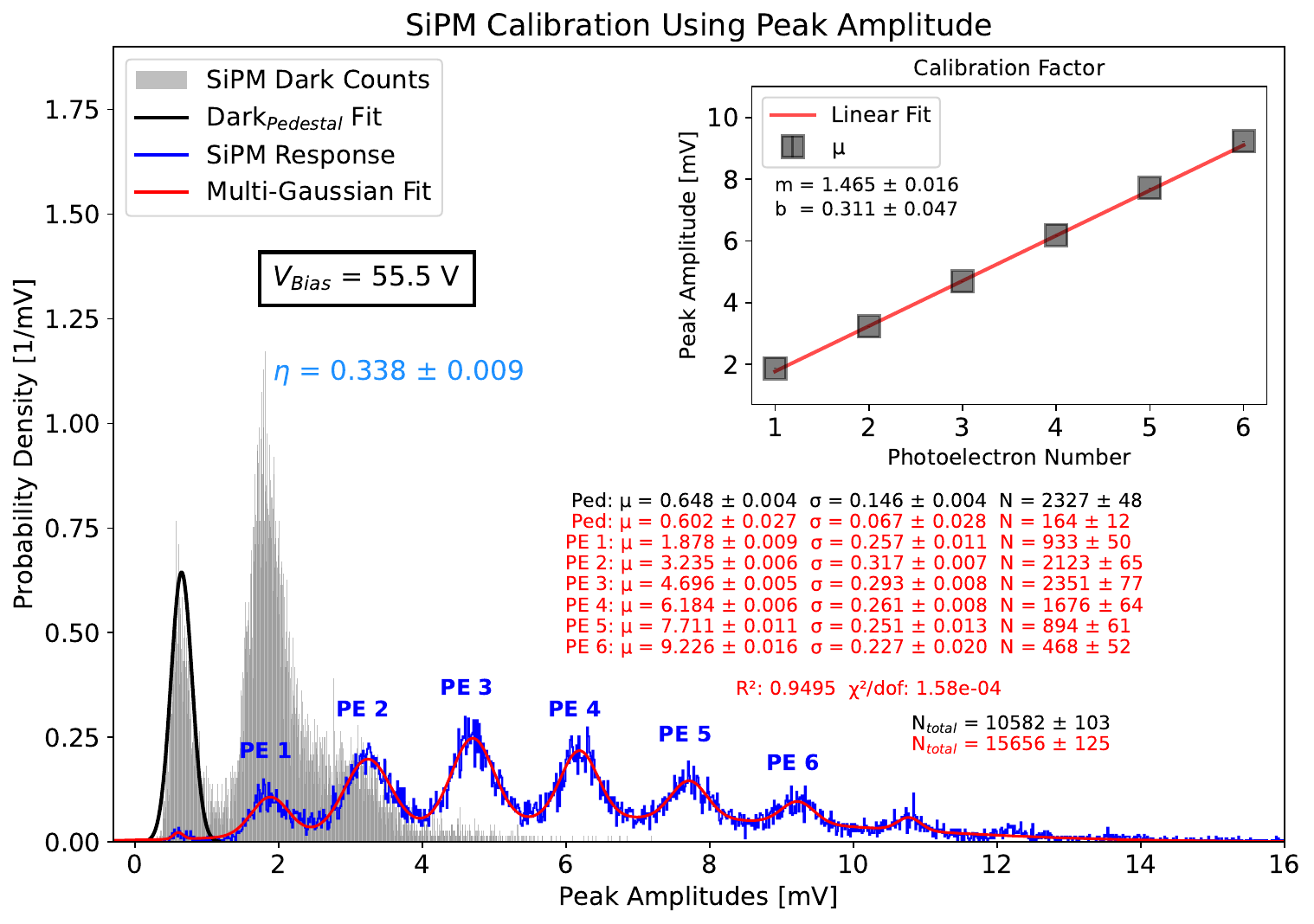}%
    \hspace{1em}%
    \includegraphics[width=0.8\textwidth]{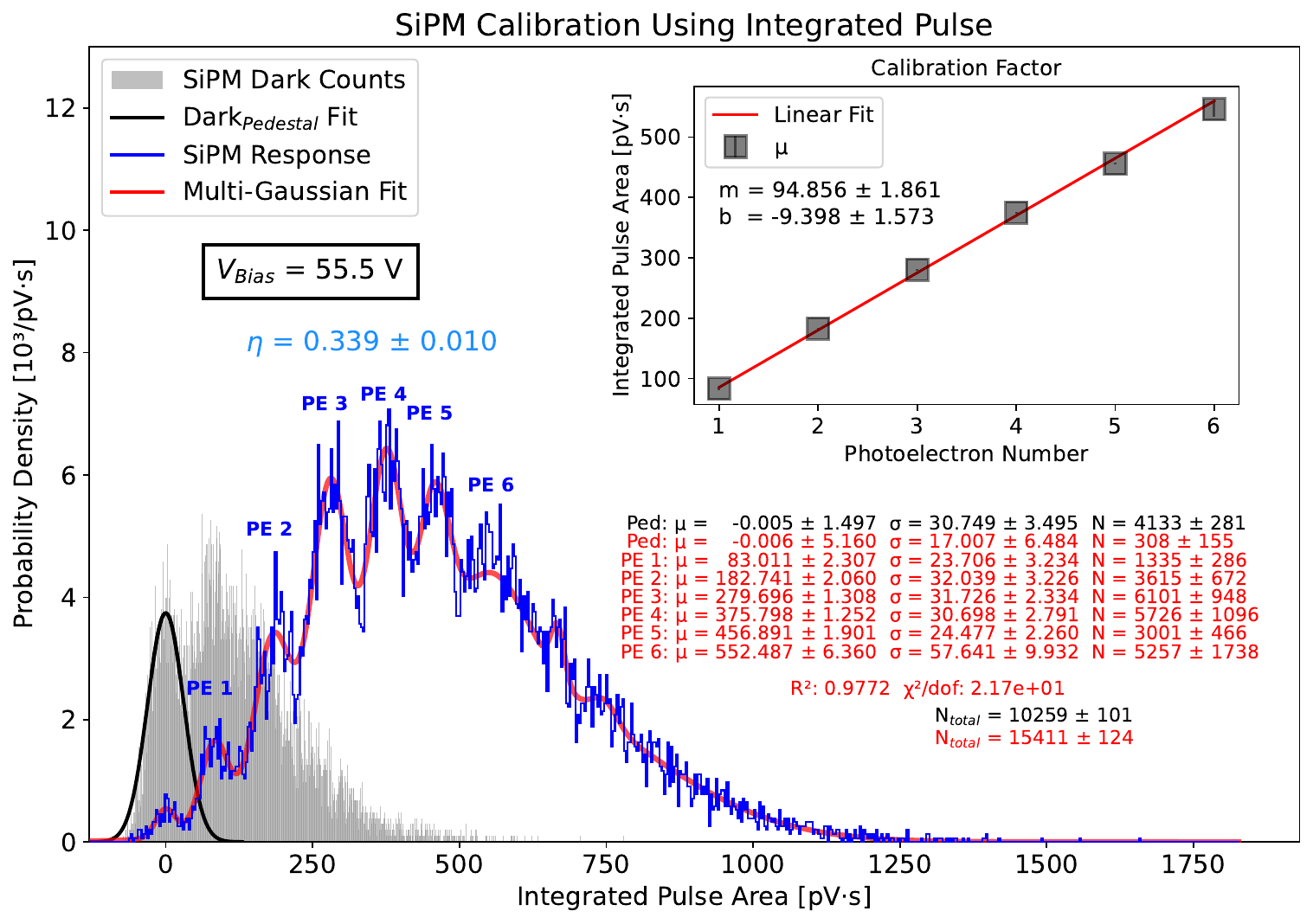}
    \caption{Comparison of SiPM photoelectron calibration using (top) peak amplitude and (bottom) integrated pulse area. Each panel shows histograms of the SiPM response (blue) with Gaussian fits to the first six photoelectron peaks and the dark pedestal (gray). The inset shows the extracted calibration factor in either mV/PE or pV·s/PE with a linear fit to the photoelectron peak positions.}
    \label{fig:CalCompare}
\end{figure}

Figure~\ref{fig:CalCompare} compares two SiPM calibration methods at room temperature and a fixed bias of 55.5\,V: one based on peak amplitude, the other on integrated pulse area. Both calibrations were derived from response histograms by fitting the first six photoelectron peaks with Gaussians. The resulting calibration constants were $1.465 \pm 0.016$\,mV/PE and $94.856 \pm 1.861$\,pV$\cdot$s/PE, respectively, with similar detection efficiencies: $\eta = 0.338 \pm 0.009$ and $\eta = 0.339 \pm 0.010$. To evaluate consistency between the two approaches, the 401\,nm picosecond laser was pulsed through each filter stack and 1500 waveforms were acquired. Integrated pulse area and peak amplitude measurements were performed on the oscilloscope, producing histograms of both quantities.

These histograms were analyzed using Gaussian fits to extract the means for pulse area $\mu_A$ and peak amplitude $\mu_P$. Table~\ref{tab:sipmcom} reports the resulting means in both physical units and photon counts calculated using the corresponding calibration factors via Equation~\ref{eq:photon_conversion}, along with the spread in photon flux $\sigma_A$ and $\sigma_P$. The uncertainty on each mean is approximately $\pm 1$ photon. The relative difference (RD) is defined as $(\mu_A - \mu_P) / [0.5(\mu_A + \mu_P)]$ and quantifies the fractional discrepancy between the two methods. Agreement between the two calibration methods is within 5\% across all filter stacks. 

\begin{table}[ht]
\centering
\begin{tabular}{c c c c c c}
\toprule
\shortstack{Filter\\Stack} & \shortstack{$\mu_A$ [nV·s] \\ (photons)} & \shortstack{$\sigma_A$ [nV·s] \\ (photons)} & \shortstack{$\mu_P$ [mV] \\ (photons)} & \shortstack{$\sigma_P$ [mV] \\ (photons)} & RD [\%] \\
\midrule
H   & 73307.8 (2280) & 1916.8 (60)  & 1136.9 (2296) & 26.0 (53)  & -0.71 \\
I   & 32097.0  (998)  & 1653.5 (51)  & 515.7 (1041)  & 25.0 (51)  & -4.25 \\
L3  & 50043.3  (1556) & 1804.5 (56)  & 775.1 (1565)  & 26.4 (53)  & -0.59 \\
L4  & 21713.9 (675)  & 1402.1 (44)  & 334.3 (675)   & 20.8 (42)  &  0.03 \\
L5  & 6550.4 (204)   & 823.3 (26)   & 101.3 (204)   & 12.0 (24)  & -0.38 \\
L6  & 17022.6 (529)  & 1325.0 (41)  & 262.0 (529)   & 19.4 (39)  &  0.04 \\
\bottomrule
\end{tabular}
\caption{Comparison of photon counts extracted using integrated pulse area $\mu_A$ and peak amplitude $\mu_P$ calibration methods for six filter stacks (see Table~\ref{tab:filter_stacks}). The uncertainty on each Gaussian mean corresponds to approximately $\pm1$ photon. The quantities $\sigma_A$ and $\sigma_P$ represent the spread in photon flux extracted from the fitted Gaussian widths. The relative difference RD is defined as $(\mu_A - \mu_P) / \left[0.5(\mu_A + \mu_P)\right]$ and is reported as a percentage.}

\label{tab:sipmcom}
\end{table}

\clearpage
\bibliographystyle{unsrtnat}

\bibliography{Bibliography}

\end{document}